\documentclass[useAMS,usenatbib,onecolumn]{mn2e}

\usepackage{amsmath, amssymb}
\usepackage[dvips]{graphicx}
\usepackage{epsfig}
\usepackage{color}
\usepackage{mathtools}

\newcommand{\rot}{\mathbf{\nabla} \times}

\newcommand{\rlight}{r_{\rm L}}

\newcommand{\aap}{A\&A}
\newcommand{\mnras}{MNRAS}

\newcommand{\apj}{ApJ}
\newcommand{\apjl}{ApJL}
\newcommand{\apjs}{ApJS}

\newcommand{\revmex}{Rev. Mex. A.A.}

\title[Multipolar electromagnetic fields]{Multipolar electromagnetic fields around neutron stars: exact vacuum solutions and related properties}

\author[J. P\'etri]{J.  P\'etri$^{1}$
\thanks{E-mail: jerome.petri@astro.unistra.fr} \\
  $^{1}$Observatoire astronomique de Strasbourg, Universit\'e de Strasbourg, CNRS, UMR 7550, 11 rue de l'universit\'e, 67000 Strasbourg, France.}

\begin{document}

\date{Accepted . Received ; in original form }

\pagerange{\pageref{firstpage}--\pageref{lastpage}} 
\pubyear{2013}

\maketitle

\label{firstpage}

\begin{abstract}
The magnetic field topology in the surrounding of neutron stars is one of the key questions in pulsar magnetospheric physics. A very extensive literature exists about the assumption of a dipolar magnetic field but very little progress has been made in attempts to include multipolar components in a self-consistent way. In this paper, we study the effect of multipolar electromagnetic fields anchored in the star. We give exact analytical solutions in closed form for any order~$l$ and apply them to the retarded point quadrupole~($l=2$), hexapole~($l=3$) and octopole~($l=4$), a generalization of the retarded point dipole~($l=1$). We also compare the Poynting flux from each multipole and show that the spin down luminosity depends on the ratio $R/\rlight$, $R$ being the neutron star radius and $\rlight$ the light-cylinder radius. Therefore the braking index also depends on $R/\rlight$. As such multipole fields possess very different topology, most importantly smaller length scales compared to the dipolar field, especially close to the neutron star, we investigate the deformation of the polar caps induced by these multipolar fields. Such fields could have a strong impact on the interpretation of the pulsed radio emission suspected to emanate from these polar caps as well as on the inferred geometry deduced from the high-energy light-curve fitting and on the magnetic field strength. Discrepancies between the two-pole caustic model and our new multipole-caustic model are emphasized with the quadrupole field. To this respect, we demonstrate that working with only a dipole field can be very misleading.
\end{abstract}


\section{Introduction}

Rotating magnetized neutron stars are at the base of several compact object classes such as pulsars, magnetars, isolated neutron stars and X-ray binaries among others. Whereas the rotation period~$P$ and its braking determined via the period derivative $\dot{P}$ are well measured and constrained by observations, the topology and strength of the magnetic field anchored in the neutron star is much less well known, with large uncertainties. The magnetic field can only be inferred by some assumptions like for instance about the spin-down luminosity induced by magnetodipole braking. Magnetic field estimates from the cyclotron line emission is a notable exception where severe constraints can be put to good accuracy \citep{1978ApJ...219L.105T}.

With no a priori knowledge about the magnetic field configuration, the most simple structure is assumed namely a dipole field. A rich literature exists on the consequences of this field around neutron stars. The dipole field is almost exclusively used to describe the magnetosphere of pulsars and magnetars. Although such field can sometimes fairly explain light-curves from the pulsed radio and high-energy emission, it sometimes completely fails to account for the polarization properties of other radio pulsars. It is hard to believe that multipoles are not anchored in the neutron star crust as such fields are already present in main sequence stars such as the Sun.

\cite{1955AnAp...18....1D} was the first to compute the electromagnetic field around a rotating dipole, taking into account the finite size of the star. His solution is often quoted to explain the magnetic dipole radiation losses from neutron stars. Nevertheless, higher multipoles can also contribute to the spin-down luminosity, to a fraction depending on the ratio $R/\rlight$ and on $B_{\rm multipole}/B_{\rm dipole}$ where $R$ is the neutron star radius and $\rlight=c/\Omega$ the light-cylinder radius with $c$ the speed of light and $\Omega=2\,\upi/P$ the star rotation rate. Effects of higher multipole moments can be drastic especially close to the neutron star surface where they could be dominant. 

However accounting for neutron star braking only from magnetodipole losses is not possible due to an observed braking index~$n$ always less than 3 which contradicts the point magnetodipole losses. Some part of the Poynting flux has to be explained by another multipole or for instance by a particle outflow like a relativistic wind of electron/positron pairs, leading to a braking index of $n=1$. Let us briefly mention some previous works about multipolar magnetic fields in neutron stars. \cite{1988MNRAS.234P..57B} discussed the consequences of a non-dipolar magnetic field on the spin-down evolution of pulsars. They took into account a non-constant in time proportionality factor for the deceleration with application to pulsars available at that time. \cite{2006A&A...450L...1C} explained the braking indexes less than three by an evolving magnetic dipolar moment or the presence of a fall-back disk. \cite{2007AdSpR..40.1491Y} investigated a two component model between the pure vacuum rotating dipole and the pure polar cap relativistic outflow assuming a constant potential drop. They found a braking index always between the two extremes given by $n=1$ and $n=3$.
From an observational point of view, radio emission of pulsars usually requires the formation of vacuum gaps. It seems that for some pulsars, the surface magnetic field needs significant multipolar components \citep{2001ApJ...550..383G}. This is especially true close to the polar caps \citep{2002A&A...388..235G, 2002A&A...388..246G}.

Investigations of the effect of magnetic multipoles on the neutron star shape and its surrounding have also been done during the last decades. \cite{1979ApJS...41...75R} computed the electromagnetic multipolar fields to deduce the electromagnetic recoil of a freshly borned neutron star. He also computed the off-centred magnetic multipoles and brought some higher order corrections to the Deutsch solution. \cite{1991ApJ...373L..69K} showed that for millisecond pulsars constrains can be put on the strength of the magnetic multipole fields. \cite{1999MNRAS.307..459M} studied the decay of multipolar magnetic components in neutron stars and did not find any significant evolution able to modify the radio pulse profile. Recently, \cite{2013MNRAS.434.1658M} looked at the dipole-quadrupole-hexapole magnetic field induced deformation of a neutron star opening up the possibility for emission of gravitational waves showing various patterns and discussed implications for the braking index of magnetars.
\cite{2002MNRAS.334..743A} investigated the consequences of the presence of multipolar fields in pulsar magnetospheres close to the surface. Special attention was drawn on pair cascades with application to the sub-beams of PSR~B0943+10 and some sub-pulse drifting phenomena. However, they mainly focused on axisymmetric configurations that do not permit pulsed emission. Pair cascade was also the main topic of \cite{2011ApJ...726L..10H} who found that already in the distorted dipole field the accelerating potential increases significantly. A similar investigation was performed by \cite{2003ARep...47..613K} with application to crust heating and electron extraction. \cite{1993ApJ...408..160A} looked at the spin up of neutron stars due to accretion and found a constraint on the multipole components in the magnetosphere, claiming that it will always be dominated by the dipole. Recently, \cite{2015A&A...573A..51B} computed the solutions to the multipole field in vacuum using two scalar fields that automatically satisfy the divergencelessness constrain for the magnetic field. Our approach with vector spherical harmonics is similar to their scalar fields although that our formalism is more handy to solve for the electromagnetic field components in vacuum. They also briefly discussed the consequences on pulsar electrodynamics and pulse shape. Let us also mention that general-relativistic multipolar electromagnetic field decomposition around a rotating neutron star has been carried out more than a decade ago by \cite{2004MNRAS.352.1161R}. They estimated the corresponding spin-down luminosity for a rotating magnetic dipole and showed it to be 2 to 6 times larger than the often quoted Newtonian limit.

\cite{1998RMxAC...7..207A} studied the evolution of pulsars according to the spin-down torque produced by a monopole (wind outflow), adding a dipole and a quadrupole contribution. \cite{1983A&A...127L...1L} even discussed the possibility to detect multipole radiation from neutron stars. \cite{1998ASPC..138..293S} studied the gamma-ray emission coming from the eight gamma-ray pulsars known at that time. Curiously they found that the quadrupole fields was as strong as the dipole field suggesting a relation between high-energy emission and magnetic topology.

Understanding the very many classes of neutron stars probably requires a deep investigation of the exact topology of the multipolar fields to unveil the physics of neutron star magnetospheres. The clue resides in the relative strengths of each multipole moment. In this paper, we study in details the properties of the first four multipole moments, namely the magneto-dipole $l=1$, the magneto-quadrupole $l=2$, the magneto-hexapole~$l=3$ and the magneto-octopole~$l=4$ losses. In~\S\ref{sec:Multipole}, we give exact close analytical expressions for the electromagnetic field of a rotating multipole and for the corresponding Poynting flux in terms of spherical Hankel functions. We then specialize our treatment to the dipole, quadrupole, hexapole and octopole in~\S\ref{sec:LowMultipole}. In the appendix~\ref{app:B} we also give simple approximations for those multipoles in the limit of a vanishing radius of the star, the so called point multipole approximation, valid to good accuracy for all stars with $R\ll\rlight$ which is the case even for millisecond pulsars (for which $R/\rlight\lesssim0.1$). The consequences of such fields are then investigated. The associated Poynting flux and braking indexes are given in~\S\ref{sec:Poynting}. Upper limits for the magnetic multipole components are given according to the magneto-multipole losses formula. The polar cap geometry is strongly influenced by the non-dipolar components as shown in~\S\ref{sec:PolarCap}. The pulsed high-energy emission within the magnetosphere, along the last closed field line (slot gap model) shows drastically different properties compared to the dipole as discussed in~\S\ref{sec:Emission}. Conclusions and ongoing work are drawn in~\S\ref{sec:Conclusion}.

\section{Exact multipolar electromagnetic fields}
\label{sec:Multipole}

Finding the multipolar electromagnetic field of a rotating magnetized sphere can be reduced to a set of homogeneous wave equations for scalar quantities associated to the electric and magnetic field. The general framework in general relativity has been given in \cite{2013MNRAS.433..986P}. Here we specialize the discussion to Minkowski space-time \citep{2012MNRAS.424..605P}, avoiding frame-dragging effects which would generate higher multipoles from lower ones. All the details about the derivation of the partial differential equations recalled below and the associated expansion method are taken from chapter~9 of \cite{2001elcl.book.....J} dealing with radiating systems and multipolar fields. However, our definitions do not strictly follow the usual conventions taken for quantum mechanics. We give exact analytical solutions for any multipole moment, specifying boundary conditions on the star being treated as a perfect conductor.

\subsection{General solution}

Let us assume that the neutron star interior is filled with a constant magnetic field dragged into solid uniform rotation by the star. The magnetic field can then be cast into an expansion according to the vector spherical harmonics such that in spherical coordinates~$(r,\vartheta,\varphi)$
\begin{equation}
 \textbf{\textit{B}}_{\rm in}(r,\vartheta,\varphi,t) = \sum_{l=1}^\infty\sum_{m=-l}^l \left( \rot [f^{\rm B}_{l,m}(r) \, \mathbf{\Phi}_{l,m} ] \, e^{-i\,m\,\Omega\,t} \right).
\end{equation}
The exponential term $e^{-i\,m\,\Omega\,t}$ emphasizes the solid body rotation of the magnetic field structure as seen by a distant observer in its inertial frame. We recall that the azimuthal term $e^{i\,m\,\varphi}$ can be factored out of $\mathbf{\Phi}_{l,m}$ such that each mode $m$ behaves as $e^{i\,m\,(\varphi - \Omega\,t)}$ corresponding indeed to corotation at the neutron star speed~$\Omega$.
For $r\leq R$, the functions $f^{\rm B}_{l,m}(r)$ are prescribed by the particular magnetic field configuration inside the star. These are known functions and will impose the boundary conditions on the neutron star surface. Note that the dependence on time is entirely contained in the harmonic terms, the complex exponentials $e^{-i\,m\,\Omega\,t}$. It is understood that the physical solution corresponds to the real part of $\textbf{\textit{B}}_{\rm in}$. The vector spherical harmonics~$\mathbf{\Phi}_{l,m}$ are defined in \cite{2013MNRAS.433..986P}. To the lowest order in the expansion, there is no monopole field therefore no $l=0$ term. We are interested in the exact multipolar electromagnetic field induced by a rotating neutron star in a flat vacuum space-time. The divergencelessness constraint on the electric field and magnetic field can be conveniently expressed in terms of vector spherical harmonics~$\mathbf{\Phi}_{l,m}$. Indeed the electromagnetic field outside the star is also developed on to vector spherical harmonics such that
\begin{subequations}
\begin{align}
  \label{eq:Decomposition_HSV_div_0_D}
  \mathbf{D}_{\rm out}(r,\vartheta,\varphi,t) = & \sum_{l=1}^\infty\sum_{m=-l}^l \left( \rot [f^{\rm D}_{l,m}(r,t) \, \mathbf{\Phi}_{l,m}] + g^{\rm D}_{l,m}(r,t) \, \mathbf{\Phi}_{l,m} \right) \\
  \label{eq:Decomposition_HSV_div_0_B}
  \mathbf{B}_{\rm out}(r,\vartheta,\varphi,t) = & \sum_{l=1}^\infty\sum_{m=-l}^l \left( \rot [f^{\rm B}_{l,m}(r,t) \, \mathbf{\Phi}_{l,m}] + g^{\rm B}_{l,m}(r,t) \, \mathbf{\Phi}_{l,m} \right).
\end{align}
We are looking for stationary solutions expressed in the observer frame. This is achieved by a temporal dependence of $f_{l,m}$ and $g_{l,m}$ given by $e^{-i\,m\,\Omega\,t}$. Thus the
functions $g^{\rm D}_{l,m}$ and $g^{\rm B}_{l,m}$ are related to the function $f^{\rm B}_{l,m}$ and  $f^{\rm D}_{l,m}$ according to Maxwell equations by a simple linear scaling such that
\begin{align}
\label{eq:gDvsfB}
 g^{\rm D}_{l,m} = & + i \, \varepsilon_0 \, m \, \Omega \, f^{\rm B}_{l,m} \\
\label{eq:gBvsfD}
 g^{\rm B}_{l,m} = & - i \,         \mu_0 \, m \, \Omega \, f^{\rm D}_{l,m}.
\end{align}
\end{subequations}
We recall that in flat space-time vacuum the constitutive relations are $\mathbf{D}=\varepsilon_0\,\mathbf{E}$ and $\mathbf{B}=\mu_0\,\mathbf{H}$.
For a general multipolar electromagnetic field, the functions $\{f^{\rm B}_{l,m},f^{\rm D}_{l,m}\}$ are separately solutions of an Helmholtz equation given by
\begin{equation}
\label{eq:Helmholtz}
 \frac{1}{r} \, \partial_r^2(r\,f_{l,m}) - \frac{l\,(l+1)}{r^2} \, f_{l,m} + k_m^2 \, f_{l,m} = 0
\end{equation}
where we introduced the wave-number $k_m=m\,k$ and
\begin{equation}
 k = \frac{\Omega}{c} = \frac{1}{\rlight}.
\end{equation}
In such a way, we decoupled the electric part $f^{\rm D}_{l,m}$ from the magnetic part $f^{\rm B}_{l,m}$. This decoupling arises from the definition of the vector spherical harmonics $\mathbf{\Phi}_{l,m}$ being eigenfunctions of the angular Laplacian differential operator in the sense that
\begin{equation}
 \Delta_{\vartheta,\varphi} \mathbf{\Phi}_{l,m} = \left[ \frac{1}{\sin\vartheta} \, \frac{\partial}{\partial \vartheta} \sin\vartheta \, \frac{\partial}{\partial \vartheta} + \frac{1}{\sin^2\vartheta} \, \frac{\partial^2}{\partial \varphi^2} \right] \mathbf{\Phi}_{l,m} = - l\,(l+1) \, \mathbf{\Phi}_{l,m}.
\end{equation}
Equation~(\ref{eq:Helmholtz}) and its general solution can be found in any textbook such as for instance in \cite{2001elcl.book.....J}.
The solution corresponding to a spherical outgoing wave, that with $m>0$, is
\begin{equation}
 f_{l,m} = a_{l,m} \, h_l^{(1)}(k_m\,r)
\end{equation}
where $h_l^{(1)}(x)$ is the spherical Hankel function of order~$l$ as defined by \cite{2005mmp..book.....A}. The constant $a_{l,m}$ is determined by the boundary conditions on the star, namely continuity of the radial component of $\mathbf{B}$ and continuity of the tangential component of $\mathbf{D}$. We will come to this point later again. Moreover the axisymmetric modes~$m=0$ do not propagate. They decay to zero at infinity according to a power law in radius given by
\begin{equation}
 f_{l,0} = a_{l,0} \, r^{-(l+1)} .
\end{equation}
$a_{l,0}$ are also constant to be determined by the boundary conditions that, on the neutron star surface, enforce continuity of the radial component of the magnetic, implying continuity of the functions $f^{\rm B}_{l,m}$ at the stellar surface. For the electric field, we must satisfy for any~$m$
\begin{equation}
 \label{eq:BC_fD}
 \partial_r ( r \, f^{\rm D}_{l,m}) = \varepsilon_0 \, r \, \Omega \, \left[ \sqrt{(l+1)\,(l-1)} \, J_{l,m} \, f^{\rm B}_{l-1,m} 
 - \sqrt{l\,(l+2)} \, J_{l+1,m} \, f^{\rm B}_{l+1,m} \right]
\end{equation}
where $J_{l,m} = \sqrt{\frac{l^2-m^2}{4\,l^2-1}}$, see \cite{2013MNRAS.433..986P}. These relations fully determine the electromagnetic field outside the star for any multipolar component.
Indeed, the general solution reads
\begin{subequations}
\begin{align}
  \label{eq:Solution_Generale_div_0_D}
  \mathbf{D}_{\rm out}(r,\vartheta,\varphi,t) = & \sum_{l=1}^\infty \rot [a^{\rm D}_{l,0} \, \frac{\mathbf{\Phi}_{l,0}}{r^{l+1}}] + \sum_{l=1}^\infty\sum_{m=-l,m\neq0}^l  \left( \rot [a^{\rm D}_{l,m} \, h_l^{(1)}(k_m\,r) \, \mathbf{\Phi}_{l,m}]  + i \, \varepsilon_0 \, m \, \Omega \, a^{\rm B}_{l,m} \, h_l^{(1)}(k_m\,r) \, \mathbf{\Phi}_{l,m} \right) \, e^{-i\,m\,\Omega\,t} \\
  \label{eq:Solution_Generale_div_0_B}
  \mathbf{B}_{\rm out}(r,\vartheta,\varphi,t) = & \sum_{l=1}^\infty \rot [a^{\rm B}_{l,0} \, \frac{\mathbf{\Phi}_{l,0}}{r^{l+1}}] + \sum_{l=1}^\infty\sum_{m=-l,m\neq0}^l  \left( \rot [a^{\rm B}_{l,m} \, h_l^{(1)}(k_m\,r) \, \mathbf{\Phi}_{l,m}]  - i \, \mu_0 \, m \, \Omega \, a^{\rm D}_{l,m} \, h_l^{(1)}(k_m\,r) \, \mathbf{\Phi}_{l,m} \right) \, e^{-i\,m\,\Omega\,t}
\end{align}
\end{subequations}
where $\{a^{\rm D}_{l,m}, a^{\rm B}_{l,m}\}$ are constants depending on the boundary conditions imposed on the stellar surface. Note that the presence of $a^{\rm D}_{l,m}$ terms in the magnetic field and of $a^{\rm B}_{l,m}$ terms in the electric field does not contradict the fact that equations~(\ref{eq:Helmholtz}) are decoupled. The coupling is due to the $g_{l,m}$ functions as defined in equations~(\ref{eq:gDvsfB}),(\ref{eq:gBvsfD}). Any particular solution to the field equation requires the determination of these constants $\{a^{\rm D}_{l,m}, a^{\rm B}_{l,m}\}$. First the continuity of the radial component of $\mathbf{B}$ for $m>0$ gives
\begin{equation}
\label{eq:aBlm}
 a^{\rm B}_{l,m} = \frac{f^{\rm B}_{l,m}(R)}{h_l^{(1)}(k_m\,R)}
\end{equation}
and for the axisymmetric cases~$m=0$
\begin{equation}
\label{eq:aBl0}
 a^{\rm B}_{l,0} = R^{l+1} \, f^{\rm B}_{l,0}(R).
\end{equation}
Second, the continuity of the tangential component of the electric field gives for the axisymmetric and non axisymmetric coefficients respectively,
\begin{subequations}
\begin{align}
 \label{eq:BC_aD}
 l \, a^{\rm D}_{l0} & = - \varepsilon_0 \, \Omega \, R^{l+2} \, \left[ \sqrt{(l+1)\,(l-1)} \, J_{l,0} \, f^{\rm B}_{l-1,0}(R) - \sqrt{l\,(l+2)} \, J_{l+1,0} \, f^{\rm B}_{l+1,0}(R) \right] \\
 a^{\rm D}_{l,m} \, \left.\partial_r ( r \, h_l^{(1)}(k_m\,r))\right|_{r=R} & = \varepsilon_0 \, R \, \Omega \, \left[ \sqrt{(l+1)\,(l-1)} \, J_{l,m} \, f^{\rm B}_{l-1,m}(R) - \sqrt{l\,(l+2)} \, J_{l+1,m} \, f^{\rm B}_{l+1,m}(R) \right] 
\end{align}
\end{subequations}
where derivatives are evaluated at the stellar surface $r=R$. This represents the most general expression to the vacuum multipolar electromagnetic field outside the neutron star. Knowing the radial component of the magnetic field on its surface, we are able to compute straightforwardly the full electromagnetic field in vacuum. In what follows, we specialize these solutions to some practical situations including low order multipoles like the magnetic dipole, quadrupole, hexapole and octopole. We also give useful and handy compact expressions in the limit of vanishing radius of the neutron star: we refer to it as retarded point multipole field, see~\S\ref{sec:LowMultipole} and the appendix~\ref{app:B}.
Before, we show how to compute the exact multipole Poynting flux for any mode $(l,m)$.
To this end, we look for the asymptotic expression of the electromagnetic field in the wave zone.

\subsection{Wave zone and Poynting flux}

In the wave zone, the above expressions can be drastically reduced by the fact that the spherical Hankel functions behave asymptotically as $h_l^{(1)}(x) \approx (-i)^{l+1} \, e^{ix}/x$ \citep{2005mmp..book.....A}. Neglecting the axisymmetric mode decreasing much faster, like $r^{-(l+1)}$, the electromagnetic field becomes in the limit $r\gg\rlight$
\begin{subequations}
\begin{align}
  \label{eq:Solution_Asymptot_B}
  \mathbf{B}_{\rm w} & = \sum_{l\geq1,m\neq0} - (-i)^{l} \, \frac{e^{i\,(k_m\,r - m\,\Omega\,t)}}{k_m\,r}
    \left( k_m \, a^{\rm B}_{l,m} \, \mathbf{\Psi}_{l,m} + \mu_0 \, m \, \Omega \, a^{\rm D}_{l,m} \, \mathbf{\Phi}_{l,m} \right) \\
  \label{eq:Solution_Asymptot_D}
  \mathbf{D}_{\rm w} & = \sum_{l\geq1,m\neq0} - (-i)^{l} \, \frac{e^{i\,(k_m\,r - m\,\Omega\,t)}}{k_m\,r}    \left( k_m \, a^{\rm D}_{l,m} \, \mathbf{\Psi}_{l,m} - \varepsilon_0 \, m \, \Omega \, a^{\rm B}_{l,m} \, \mathbf{\Phi}_{l,m} \right)  \\
  \label{eq:OndePlane}
  = & \varepsilon_0 \, c \, \mathbf{B}_{\rm w} \wedge \mathbf{n}.
\end{align}
\end{subequations}
Equation~(\ref{eq:OndePlane}) shows that the solution behaves as a monochromatic plane wave propagating in the radial direction $\mathbf{n} = \mathbf{e}_{\rm r}$ at frequency~$\Omega$. It is derived from the identities
\begin{subequations}
 \begin{align}
  \mathbf{n} \wedge \mathbf{\Psi}_{l,m} & = \mathbf{\Phi}_{l,m} \\
  \mathbf{n} \wedge \mathbf{\Phi}_{l,m} & = - \mathbf{\Psi}_{l,m} .
 \end{align}
\end{subequations}
The time averaged Poynting flux is therefore
\begin{equation}
 \mathbf{S} = \frac{\mathbf{D}_{\rm w} \wedge \mathbf{B}_{\rm w}^*}{2\,\mu_0\,\varepsilon_0}
\end{equation}
where $\mathbf{B}_{\rm w}^*$ is the complex conjugate of $\mathbf{B}_{\rm w}$. Integrating the radial component of the Poynting vector along the solid angle we get the power radiated, using the orthonormality of the vector spherical harmonics, such that
\begin{equation}
\label{eq:luminosite_multipole}
 L = \int_\Omega S_{\rm r} \, r^2 \, d\Omega = \frac{c}{2\,\mu_0} \, \sum_{l\geq1,m\neq0} \left( |a^{\rm B}_{l,m}|^2 + \mu_0^2 \, c^2 \, |a^{\rm D}_{l,m}|^2 \right) .
\end{equation}
The spin down luminosity~$L$ is independent of the radius as it should from the energy conservation law. Equation~(\ref{eq:luminosite_multipole}) represents the most general expression for the magneto-multipole losses from an arbitrary multipole magnetic field, see for instance also \cite{2001elcl.book.....J}.

\subsection{Exact solution for one multipole}

It is instructive to find the exact solution for a particular multipole field with fixed numbers $(l,m)$. Let us assume that inside the star, the magnetic field is solely represented by the function $f^{\rm B}_{l,m}(r)$. What should then the electromagnetic field be outside the star? The only non vanishing magnetic field coefficient is given by equation~(\ref{eq:aBlm}) or equation~(\ref{eq:aBl0}) if $m=0$. Moreover the two non-vanishing electric field coefficients are (if $l=1$ only one solution exists, see the dipole case below) for $m>0$
\begin{subequations}
\label{eq:aDlm}
\begin{align}
 a^{\rm D}_{l+1,m} \, \left.\partial_r ( r \, h_{l+1}^{(1)}(k_m\,r))\right|_{r=R} & = \varepsilon_0 \, R \, \Omega \,  \sqrt{l\,(l+2)} \, J_{l+1,m} \, f^{\rm B}_{l,m}(R) \\
 a^{\rm D}_{l-1,m} \, \left.\partial_r ( r \, h_{l-1}^{(1)}(k_m\,r))\right|_{r=R} & = - \varepsilon_0 \, R \, \Omega \, \sqrt{(l-1)\,(l+1)} \, J_{l,m} \, f^{\rm B}_{l,m}(R)
 \end{align}
\end{subequations}
and for the axisymmetric case we find
\begin{subequations}
\begin{align}
\label{eq:aDl0}
 (l+1) \, a^{\rm D}_{l+1,0} & = - \varepsilon_0 \, R^{l+3} \, \Omega \,  \sqrt{l\,(l+2)} \, J_{l+1,0} \, f^{\rm B}_{l,0}(R) \\
 (l-1) \, a^{\rm D}_{l-1,0} & =   \varepsilon_0 \, R^{l+1} \, \Omega \, \sqrt{(l-1)\,(l+1)} \, J_{l,0} \, f^{\rm B}_{l,0}(R).
 \end{align}
\end{subequations}
We conclude that the solution is fully specified by the three constants of integration $(a^{\rm B}_{l,m}, a^{\rm D}_{l+1,m}, a^{\rm D}_{l-1,m})$. The Poynting flux associated to this particular solution is for $m>0$ (it vanishes for $m=0$)
\begin{subequations}
\label{eq:luminosite_un_multipole}
\begin{align}
 P_{l,m} & = \frac{c}{2\,\mu_0} \, \left[ |a^{\rm B}_{l,m}|^2 + \mu_0^2 \, c^2 \, ( |a^{\rm D}_{l-1,m}|^2 + |a^{\rm D}_{l+1,m}|^2 ) \right] \\
 & = \frac{c|f^{\rm B}_{l,m}(R)|^2}{2\,\mu_0} \, \mathcal{S}_{l,m} \\
 \mathcal{S}_{l,m} & = \frac{1}{|h_l^{(1)}(k_m\,R)|^2} + \frac{R^2}{\rlight^2} \, \left( \frac{l(l+2)\,J^2_{l+1,m}}{|\partial_r ( r \, h_{l+1}^{(1)}(k_m\,r))|^2_R} + \frac{(l-1)(l+1)\,J^2_{l,m}}{|\partial_r ( r \, h_{l-1}^{(1)}(k_m\,r))|^2_R} \right) .
 \end{align}
\end{subequations}
For the special case $l=1$, the constant $a^{\rm D}_{l-1,m}$ does not exist.
For later purposes, it will be useful to get the series expansions to first order of $\mathcal{S}_{l,m}$ in terms of $a=R/\rlight$. For completeness, we give them in Table~\ref{tab:Series}. Actually they represent within a multiplicative factor the Poynting flux for the retarded point multipole field.
\begin{table}
 \centering
 \begin{center}
\begin{tabular}{cccccc}
 \hline
 $l/m$ & 1 & 2 & 3 & 4 & Normalization \\
 \hline
 \hline
 1 & $1$ &  &  & & $a^4$ \\
 2 & $\frac{32}{45}$ & $\frac{64}{9}$ & & & $a^6$ \\
 3 & $\frac{29}{525}$ & $\frac{1664}{525}$ & $\frac{729}{25}$ & & $a^8$ \\
 4 & $\frac{184}{99225}$ & $\frac{45056}{99225}$ & $\frac{13176}{1225}$ & $\frac{1048576}{11025}$ & $a^{10}$ \\
 \hline
 \end{tabular}
 \end{center}
 \caption{First order expansion of the function $\mathcal{S}_{l,m}$ in terms of $a=R/\rlight$. The first column indicates the multipole order~$l$ and the first line indicates the azimuthal mode~$m$ with the restriction $m\leqslant l$. The normalization is proportional to $a^{2\,l+2}\propto\Omega^{2\,l+2}$, typical for a $l$-multipole.}
 \label{tab:Series}
\end{table}
For concreteness, we switch now to explicit application of low order multipole solutions in order to get more physical insight into their properties. For the remainder of this paper, we focus on some illuminating cases such as the low order multipoles: dipole, quadrupole, hexapole and octopole. To end this general discussion about the computation of multipolar fields, we compare our results with the recent work done by \cite{2015A&A...573A..51B}. It is shown in the appendix~\ref{app:A} that we get exactly the same expressions as \cite{2015A&A...573A..51B} except for an error in the sign of their electric field for the axisymmetric multipoles.

\section{Explicit low multipole solutions}
\label{sec:LowMultipole}

Multipolar magnetic fields are important not only for neutron stars but also for main sequence stars, or any other star possessing a non negligible magnetic field. To study quantitatively the influence of those components, it is illuminating to get simple closed analytical expressions for those fields. Thus in this section, we give the full exact solutions for the dipole, quadrupole, hexapole and octopole fields. For practical applications, we also give the approximation for a retarded point multipole useful and very accurate for slowly rotating stars as those we study here: by slowly rotating we mean $R\ll\rlight$.

\subsection{The magnetic dipole $l=1$}

We start our discussion with the well known magnetic dipole. The ease and power of our description is demonstrated by application to the dipole field. Introducing the vector spherical harmonics expansion and two free parameters $\{q_{1,0}, q_{1,1}\}$ related to the strength of each azimuthal mode~$m$, the \textit{quasi-static} rotating dipole frozen into the neutron star is conveniently written as
\begin{equation}
\label{eq:Bdipole}
 \textit{\textbf{B}} = B \, R^3 \, \textrm{Re} \left\lbrace \rot \left[ q_{1,0} \, \frac{\mathbf{\Phi_{1,0}}}{r^2} + q_{1,1} \, \frac{\mathbf{\Phi_{1,1}}}{r^2} \, e^{-i\,\Omega\,t} \right] \right\rbrace .
\end{equation}
The symbol $\textrm{Re}$ means the real part and the two constant coefficients $\{q_{1,0}, q_{1,1}\}$ are traditionally expressed in terms of the magnetic obliquity~$\chi$, angle between magnetic dipole vector and rotation axis, such that
\begin{subequations}
 \begin{align}
  q_{1,0} & = -\sqrt{\frac{8\,\upi}{3}} \, \cos\chi \\
  q_{1,1} & = \sqrt{\frac{16\,\upi}{3}} \, \sin\chi.
 \end{align}
\end{subequations}
Thus $\chi=0^o$ gives the axisymmetric mode $m=0$ alone whereas $\chi=90^o$ give the $m=1$ mode alone. Expression~(\ref{eq:Bdipole}) is only valid close to the star where retardation effects are negligible. It corresponds to the quasi-static limit, the lowest order approximation in $R/\rlight$. Actually it also represents the boundary value imposed by the rotating neutron star. Solving the Helmholtz equation~(\ref{eq:Helmholtz}) for the unknown expansion functions~$\{f^{\rm D}_{l,m}, f^{\rm B}_{l,m}\}$, the rotating magnetic dipole inside the star is advantageously expressed by only two non-vanishing functions given by
\begin{subequations}
\begin{align}
 f^{\rm B}_{1,0} & = - \sqrt{\frac{8\,\pi}{3}} \, \frac{B \, R^3 \, \cos\chi}{r^2} \\
 f^{\rm B}_{1,1} & = \sqrt{\frac{16\,\pi}{3}} \, \frac{B \, R^3 \, \sin\chi}{r^2}.
\end{align}
\end{subequations}
The boundary conditions impose continuity of $\{f^{\rm B}_{1,0},f^{\rm B}_{1,1}\}$ at the stellar surface and for the electric field we get
\begin{subequations}
 \begin{align}
 \partial_r(r\,f_{2,0}^{\rm D}) & = \frac{2}{\sqrt{5}} \, \varepsilon_0 \, R \, \Omega \, f_{1,0}^{\rm B} \\
 \partial_r(r\,f_{2,1}^{\rm D}) & = \sqrt{\frac{3}{5}} \, \varepsilon_0 \, R \, \Omega \, f_{1,1}^{\rm B}.
 \end{align}
\end{subequations}
The full solution in vacuum outside the star therefore becomes
\begin{subequations}
\label{eq:Deutsch}
 \begin{align}
  a_{1,0}^{\rm B} & = -\sqrt{\frac{8\,\upi}{3}} \, B \, R^3 \, \cos\chi \\
  a_{1,1}^{\rm B} & = \sqrt{\frac{16\,\upi}{3}} \, \frac{B \, R \, \sin\chi}{h_1^{(1)}(k\,R)} \\
  a_{2,0}^{\rm D} & = \sqrt{\frac{8\,\upi}{15}} \, \varepsilon_0 \, \Omega \, B \, R^5 \, \cos\chi \\
  a_{2,1}^{\rm D} & = \sqrt{\frac{16\,\upi}{5}} \, \varepsilon_0 \, \frac{\Omega \, B \, R^2 \, \sin\chi}{\left. \partial_r(r\,h_2^{(1)}(k\,r))\right|_{r=R}}
 \end{align}
\end{subequations}
which is nothing else than Deutsch solution \citep{1955AnAp...18....1D}. It is easy to check that this configuration with only one free parameter corresponding to the inclination angle of the magnetic moment~$\chi$ leaves the total magnetic energy outside the star constant. Indeed the total magnetic energy amounts to
\begin{equation}
 W_{\rm mag } = \int\int\int \frac{B^2}{2\,\mu_0} \, dV = \frac{B^2 \, R^3}{4\,\mu_0} \, (2 \, q_{1,0}^2 + q_{1,1}^2 ) = 4\,\upi\,\frac{B^2 \, R^3}{3\,\mu_0}
\end{equation}
independently of~$\chi$. $dV$ stands for the integration in the whole three dimensional space outside the star. The total magnetic energy in vacuum outside the star will be used to scale the different multipole moments with respect to the dipole one.

\subsection{The magnetic quadrupole $l=2$}

Let us perform the same expansion to the magnetic quadrupole such that it can be expressed inside the star by
\begin{equation}
 \textit{\textbf{B}} = B \, R^4 \, \textrm{Re} \left\lbrace \rot \left[ q_{2,0} \, \frac{\mathbf{\Phi_{2,0}}}{r^3} + q_{2,1} \, \frac{\mathbf{\Phi_{2,1}}}{r^3} \, e^{-i\,\Omega\,t} + q_{2,2} \, \frac{\mathbf{\Phi_{2,2}}}{r^3} \, e^{-2\,i\,\Omega\,t} \right] \right\rbrace 
\end{equation}
The boundary conditions impose continuity of $\{f^{\rm B}_{2,0},f^{\rm B}_{2,1},f^{\rm B}_{2,2}\}$ at the stellar surface and for the tangential component of the electric field we get
\begin{subequations}
 \begin{align}
 \partial_r(r\,f_{1,0}^{\rm D}) & = - \frac{2}{\sqrt{5}} \, \varepsilon_0 \, R \, \Omega \, f_{2,0}^{\rm B} \\
 \partial_r(r\,f_{1,1}^{\rm D}) & = - \sqrt{\frac{3}{5}} \, \varepsilon_0 \, R \, \Omega \, f_{2,1}^{\rm B} \\
 \partial_r(r\,f_{3,0}^{\rm D}) & = 6 \, \sqrt{\frac{2}{35}} \, \varepsilon_0 \, R \, \Omega \, f_{2,0}^{\rm B} \\
 \partial_r(r\,f_{3,1}^{\rm D}) & = \frac{8}{\sqrt{35}} \, \varepsilon_0 \, R \, \Omega \, f_{2,1}^{\rm B} \\
 \partial_r(r\,f_{3,2}^{\rm D}) & = 2\,\sqrt{\frac{2}{7}} \, \varepsilon_0 \, R \, \Omega \, f_{2,2}^{\rm B} .
 \end{align}
\end{subequations}
Solving again Helmholtz equation~(\ref{eq:Helmholtz}), the solution in vacuum is represented by
\begin{subequations}
 \begin{align}
  a_{2,0}^{\rm B} & = q_{2,0} \, B \, R^4 \\
  a_{2,1}^{\rm B} & = q_{2,1} \, \frac{B \, R}{h_2^{(1)}(k_1\,R)} \\
  a_{2,2}^{\rm B} & = q_{2,2} \, \frac{B \, R}{h_2^{(1)}(k_2\,R)} \\
  a_{1,0}^{\rm D} & = \frac{2}{\sqrt{5}} \, \varepsilon_0 \, \Omega \, B \, R^4 \, q_{2,0} \\
  a_{1,1}^{\rm D} & = - \sqrt{\frac{3}{5}} \, \varepsilon_0 \, \frac{\Omega \, B \, R^2 \, q_{2,1}}{\left. \partial_r(r\,h_1^{(1)}(k_1\,r))\right|_{r=R}} \\
  a_{3,0}^{\rm D} & = -2\,\sqrt{\frac{2}{35}} \, \varepsilon_0 \, \Omega \, B \, R^6 \, q_{2,0} \\
  a_{3,1}^{\rm D} & = \frac{8}{\sqrt{35}} \, \varepsilon_0 \, \frac{\Omega \, B \, R^2 \, q_{2,1}}{\left. \partial_r(r\,h_3^{(1)}(k_1\,r))\right|_{r=R}} \\
  a_{3,2}^{\rm D} & = 2\,\sqrt{\frac{2}{7}} \, \varepsilon_0 \, \frac{\Omega \, B \, R^2 \, q_{2,2}}{\left. \partial_r(r\,h_3^{(1)}(k_2\,r))\right|_{r=R}} .
 \end{align}
\end{subequations}
We have to fix the three constants $\{q_{2,0}, q_{2,1}, q_{2,2}\}$ in a judicious way to compare results with the dipole. To do this, we compute the total magnetic energy outside the star from its quadrupole moment and find
\begin{equation}
 W_{\rm mag } = \int\int\int \frac{B^2}{2\,\mu_0} \, dV = \frac{B^2 \, R^3}{2\,\mu_0} \, (2 \, q_{2,0}^2 + q_{2,1}^2 + q_{2,2}^2 ).
\end{equation}
Assuming a constant magnetic energy, irrespective of its geometry, the coefficients $\{q_{2,0}, q_{2,1}, q_{2,2}\}$  can be related by
\begin{subequations}
 \begin{align}
  q_{2,0} & = \sqrt{\frac{4\,\upi}{3}} \, \cos \chi_1 \\
  q_{2,1} & = \sqrt{\frac{8\,\upi}{3}} \, \sin \chi_1 \, \cos \chi_2 \\
  q_{2,2} & = \sqrt{\frac{8\,\upi}{3}} \, \sin \chi_1 \, \sin \chi_2.
 \end{align}
\end{subequations}
where $\chi_1\in[0,\upi]$ and $\chi_2\in[0,2\,\upi]$ are two angles specifying the particular geometry of the quadrupole magnetic field. With this normalization of $\{q_{2,0}, q_{2,1}, q_{2,2}\}$, the magnetic energy is the same as for the dipolar field. The different modes are easily separated by taking $\chi_1=0^o$ for $m=0$ whatever $\chi_2$, $(\chi_1,\chi_2)=(90^o,0^o)$ for $m=1$ and $(\chi_1,\chi_2)=(90^o,90^o)$ for $m=2$.

\subsection{The magnetic hexapole $l=3$}

We show how our approach is straightforwardly extended to hexapole or even higher multipolar fields. Indeed we start with the magnetic hexapole existing inside the neutron star such that
\begin{equation}
 \textit{\textbf{B}} = B \, R^5 \, \textrm{Re} \left\lbrace \rot \left[ q_{3,0} \, \frac{\mathbf{\Phi_{3,0}}}{r^4} + q_{3,1} \, \frac{\mathbf{\Phi_{3,1}}}{r^4} \, e^{-i\,\Omega\,t} + q_{3,2} \, \frac{\mathbf{\Phi_{3,2}}}{r^4} \, e^{-2\,i\,\Omega\,t} + q_{3,3} \, \frac{\mathbf{\Phi_{3,3}}}{r^4} \, e^{-3\,i\,\Omega\,t} \right] \right\rbrace 
\end{equation}
The boundary conditions enforces continuity of $\{f^{\rm B}_{3,0}, f^{\rm B}_{3,1}, f^{\rm B}_{3,2}, f^{\rm B}_{3,3}\}$ at the stellar surface and
\begin{subequations}
 \begin{align}
 \partial_r(r\,f_{2,0}^{\rm D}) & = - 6 \, \sqrt{\frac{2}{35}} \, \varepsilon_0 \, R \, \Omega \, f_{3,0}^{\rm B} \\
 \partial_r(r\,f_{2,1}^{\rm D}) & = - \frac{8}{\sqrt{35}} \, \varepsilon_0 \, R \, \Omega \, f_{3,1}^{\rm B} \\
 \partial_r(r\,f_{2,2}^{\rm D}) & = - 2\,\sqrt{\frac{2}{7}}\, \varepsilon_0 \, R \, \Omega \, f_{3,2}^{\rm B} \\
 \partial_r(r\,f_{4,0}^{\rm D}) & = 4 \, \sqrt{\frac{5}{21}} \, \varepsilon_0 \, R \, \Omega \, f_{3,0}^{\rm B} \\
 \partial_r(r\,f_{4,1}^{\rm D}) & = \frac{5}{\sqrt{7}} \, \varepsilon_0 \, R \, \Omega \, f_{3,1}^{\rm B} \\
 \partial_r(r\,f_{4,2}^{\rm D}) & = 2 \, \sqrt{\frac{5}{7}} \, \varepsilon_0 \, R \, \Omega \, f_{3,2}^{\rm B} \\
 \partial_r(r\,f_{4,3}^{\rm D}) & = \sqrt{\frac{5}{3}} \, \varepsilon_0 \, R \, \Omega \, f_{3,3}^{\rm B}
 \end{align}
\end{subequations}
Computing the exact solutions to the Helmholtz equations (\ref{eq:Helmholtz}) leads to
\begin{subequations}
 \begin{align}
  a_{3,0}^{\rm B} & = q_{3,0} \, B \, R^5 \\
  a_{3,1}^{\rm B} & = q_{3,1} \, \frac{B \, R}{h_3^{(1)}(k_1\,R)} \\
  a_{3,2}^{\rm B} & = q_{3,2} \, \frac{B \, R}{h_3^{(1)}(k_2\,R)} \\
  a_{3,3}^{\rm B} & = q_{3,3} \, \frac{B \, R}{h_3^{(1)}(k_3\,R)} \\
  a_{2,0}^{\rm D} & = 3\,\sqrt{\frac{2}{35}} \, \varepsilon_0 \, \Omega \, B \, R^5 \, q_{3,0} \\
  a_{2,1}^{\rm D} & = - \frac{8}{\sqrt{35}} \, \varepsilon_0 \, \frac{\Omega \, B \, R^2 \, q_{3,1}}{\left. \partial_r(r\,h_2^{(1)}(k_1\,r))\right|_{r=R}} \\
  a_{2,2}^{\rm D} & = - 2\,\sqrt{\frac{2}{7}} \, \varepsilon_0 \, \frac{\Omega \, B \, R^2 \, q_{3,2}}{\left. \partial_r(r\,h_2^{(1)}(k_2\,r))\right|_{r=R}} \\
  a_{4,0}^{\rm D} & = -\sqrt{\frac{5}{21}} \, \varepsilon_0 \, \Omega \, B \, R^7 \, q_{3,0} \\
  a_{4,1}^{\rm D} & = \frac{5}{\sqrt{7}} \, \varepsilon_0 \, \frac{\Omega \, B \, R^2 \, q_{3,1}}{\left. \partial_r(r\,h_4^{(1)}(k_1\,r))\right|_{r=R}} \\
  a_{4,2}^{\rm D} & = 2\,\sqrt{\frac{5}{7}} \, \varepsilon_0 \, \frac{\Omega \, B \, R^2 \, q_{3,2}}{\left. \partial_r(r\,h_4^{(1)}(k_2\,r))\right|_{r=R}} \\
  a_{4,3}^{\rm D} & = \sqrt{\frac{5}{3}} \, \varepsilon_0 \, \frac{\Omega \, B \, R^2 \, q_{3,3}}{\left. \partial_r(r\,h_4^{(1)}(k_3\,r))\right|_{r=R}}
 \end{align}
\end{subequations}
To compare with the dipole field, we compute the total magnetic energy as
\begin{equation}
 W_{\rm mag } = \int\int\int \frac{B^2}{2\,\mu_0} \, dV = \frac{3\,B^2 \, R^3}{4\,\mu_0} \, (2 \, q_{3,0}^2 + q_{3,1}^2 + q_{3,2}^2 + q_{3,3}^2 )
\end{equation}
Assuming a constant magnetic energy in the field, irrespective of the geometry, the coefficients are related by
\begin{subequations}
 \begin{align}
  q_{3,0} & = \frac{2\,\sqrt{2\,\upi}}{3} \, \cos \chi_1 \\
  q_{3,1} & = \frac{4\,\sqrt{\upi}}{3} \, \sin \chi_1 \, \cos \chi_2 \\
  q_{3,2} & = \frac{4\,\sqrt{\upi}}{3} \, \sin \chi_1 \, \sin \chi_2 \, \cos \chi_3 \\
  q_{3,3} & = \frac{4\,\sqrt{\upi}}{3} \, \sin \chi_1 \, \sin \chi_2 \, \sin \chi_3
 \end{align}
\end{subequations}
where $\chi_1,\chi_2\in[0,\upi]$ and $\chi_3\in[0,2\,\upi]$ are three angles specifying the particular geometry of the hexapole magnetic field. With this particular normalization, the magnetic energy is the same as for the dipolar field. The different modes are easily separated by taking $\chi_1=0^o$ for $m=0$ whatever $(\chi_2,\chi_3)$, $(\chi_1,\chi_2)=(90^o,0^o)$ for $m=1$ whatever $\chi_3$, $(\chi_1,\chi_2,\chi_3)=(90^o,90^o,0^o)$ for $m=2$ and $(\chi_1,\chi_2,\chi_3)=(90^o,90^o,90^o)$ for $m=3$.

\subsection{The magnetic octopole $l=4$}

To finish this discussion about the low order multipoles, we give the exact solution to the octopole field. Inside the neutron star, the magnetic field is described a priori by a general octopolar expansion such that
\begin{equation}
 \textit{\textbf{B}} = B \, R^6 \, \textrm{Re} \left\lbrace \rot \left[ q_{4,0} \, \frac{\mathbf{\Phi_{4,0}}}{r^5} + q_{4,1} \, \frac{\mathbf{\Phi_{4,1}}}{r^5} \, e^{-i\,\Omega\,t} + q_{4,2} \, \frac{\mathbf{\Phi_{4,2}}}{r^5} \, e^{-2\,i\,\Omega\,t} + q_{4,3} \, \frac{\mathbf{\Phi_{4,3}}}{r^5} \, e^{-3\,i\,\Omega\,t} + q_{4,4} \, \frac{\mathbf{\Phi_{4,4}}}{r^5} \, e^{-4\,i\,\Omega\,t} \right] \right\rbrace 
\end{equation}
The boundary conditions for the non vanishing electric field components are listed below,
\begin{subequations}
 \begin{align}
 \partial_r(r\,f_{3,0}^{\rm D}) & = - 4 \, \sqrt{\frac{5}{21}} \, \varepsilon_0 \, R \, \Omega \, f_{4,0}^{\rm B} \\
 \partial_r(r\,f_{3,1}^{\rm D}) & = - \frac{5}{\sqrt{7}} \, \varepsilon_0 \, R \, \Omega \, f_{4,1}^{\rm B} \\
 \partial_r(r\,f_{3,2}^{\rm D}) & = - 2\,\sqrt{\frac{5}{7}}\, \varepsilon_0 \, R \, \Omega \, f_{4,2}^{\rm B} \\
 \partial_r(r\,f_{3,3}^{\rm D}) & = - \sqrt{\frac{5}{3}}\, \varepsilon_0 \, R \, \Omega \, f_{4,3}^{\rm B} \\
 \partial_r(r\,f_{5,0}^{\rm D}) & = 10 \, \sqrt{\frac{2}{33}} \, \varepsilon_0 \, R \, \Omega \, f_{4,0}^{\rm B} \\
 \partial_r(r\,f_{5,1}^{\rm D}) & = \frac{8}{\sqrt{11}} \, \varepsilon_0 \, R \, \Omega \, f_{4,1}^{\rm B} \\
 \partial_r(r\,f_{5,2}^{\rm D}) & = 2 \, \sqrt{\frac{14}{11}} \, \varepsilon_0 \, R \, \Omega \, f_{4,2}^{\rm B} \\
 \partial_r(r\,f_{5,3}^{\rm D}) & = 8 \, \sqrt{\frac{2}{33}} \, \varepsilon_0 \, R \, \Omega \, f_{4,3}^{\rm B} \\
 \partial_r(r\,f_{5,4}^{\rm D}) & = 2 \, \sqrt{\frac{6}{11}} \, \varepsilon_0 \, R \, \Omega \, f_{4,4}^{\rm B}.
 \end{align}
\end{subequations}
Solution to the Helmholtz equations (\ref{eq:Helmholtz}) therefore leads to the complete solution in the form of its constant of integration given by
\begin{subequations}
 \begin{align}
  a_{4,0}^{\rm B} & = q_{4,0} \, B \, R^6 \\
  a_{4,1}^{\rm B} & = q_{4,1} \, \frac{B \, R}{h_4^{(1)}(k_1\,R)} \\
  a_{4,2}^{\rm B} & = q_{4,2} \, \frac{B \, R}{h_4^{(1)}(k_2\,R)} \\
  a_{4,3}^{\rm B} & = q_{4,3} \, \frac{B \, R}{h_4^{(1)}(k_3\,R)} \\
  a_{4,4}^{\rm B} & = q_{4,4} \, \frac{B \, R}{h_4^{(1)}(k_4\,R)} \\
  a_{3,0}^{\rm D} & = \frac{4}{3} \,\sqrt{\frac{5}{21}} \, \varepsilon_0 \, \Omega \, B \, R^6 \, q_{4,0} \\
  a_{3,1}^{\rm D} & = - \frac{5}{\sqrt{7}} \, \varepsilon_0 \, \frac{\Omega \, B \, R^2 \, q_{4,1}}{\left. \partial_r(r\,h_3^{(1)}(k_1\,r))\right|_{r=R}} \\
  a_{3,2}^{\rm D} & = - 2\,\sqrt{\frac{5}{7}} \, \varepsilon_0 \, \frac{\Omega \, B \, R^2 \, q_{4,2}}{\left. \partial_r(r\,h_3^{(1)}(k_2\,r))\right|_{r=R}} \\
  a_{3,3}^{\rm D} & = - \sqrt{\frac{5}{3}} \, \varepsilon_0 \, \frac{\Omega \, B \, R^2 \, q_{4,3}}{\left. \partial_r(r\,h_3^{(1)}(k_3\,r))\right|_{r=R}} \\
  a_{5,0}^{\rm D} & = -2 \, \sqrt{\frac{2}{33}} \, \varepsilon_0 \, \Omega \, B \, R^8 \, q_{4,0} \\
  a_{5,1}^{\rm D} & = \frac{8}{\sqrt{11}} \, \varepsilon_0 \, \frac{\Omega \, B \, R^2 \, q_{4,1}}{\left. \partial_r(r\,h_5^{(1)}(k_1\,r))\right|_{r=R}} \\
  a_{5,2}^{\rm D} & = 2 \, \sqrt{\frac{14}{11}} \, \varepsilon_0 \, \frac{\Omega \, B \, R^2 \, q_{4,2}}{\left. \partial_r(r\,h_5^{(1)}(k_2\,r))\right|_{r=R}} \\
  a_{5,3}^{\rm D} & = 8 \, \sqrt{\frac{2}{33}} \, \varepsilon_0 \, \frac{\Omega \, B \, R^2 \, q_{4,3}}{\left. \partial_r(r\,h_5^{(1)}(k_3\,r))\right|_{r=R}} \\
  a_{5,4}^{\rm D} & = 2 \, \sqrt{\frac{6}{11}} \, \varepsilon_0 \, \frac{\Omega \, B \, R^2 \, q_{4,4}}{\left. \partial_r(r\,h_5^{(1)}(k_4\,r))\right|_{r=R}}.
 \end{align}
\end{subequations}
The total magnetic energy outside the star is is
\begin{equation}
 W_{\rm mag } = \int\int\int \frac{B^2}{2\,\mu_0} \, dV = \frac{B^2 \, R^3}{\mu_0} \, (2 \, q_{4,0}^2 + q_{4,1}^2 + q_{4,2}^2 + q_{4,3}^2 + q_{4,4}^2 )
\end{equation}
Assuming a constant magnetic energy in the field, irrespective of the geometry, the coefficients are related by
\begin{subequations}
 \begin{align}
  q_{4,0} & = \sqrt{\frac{2\,\upi}{3}} \, \cos \chi_1 \\
  q_{4,1} & = \sqrt{\frac{4\,\upi}{3}} \, \sin \chi_1 \, \cos \chi_2 \\
  q_{4,2} & = \sqrt{\frac{4\,\upi}{3}} \, \sin \chi_1 \, \sin \chi_2 \, \cos \chi_3 \\
  q_{4,3} & = \sqrt{\frac{4\,\upi}{3}} \, \sin \chi_1 \, \sin \chi_2 \, \sin \chi_3 \, \cos \chi_4 \\
  q_{4,4} & = \sqrt{\frac{4\,\upi}{3}} \, \sin \chi_1 \, \sin \chi_2 \, \sin \chi_3 \, \sin \chi_4
 \end{align}
\end{subequations}
where $\chi_1,\chi_2,\chi_3\in[0,\upi]$ and $\chi_4\in[0,2\,\upi]$ are four angles specifying the particular geometry of the octopole magnetic field. With this particular normalization, the magnetic energy is the same as for the dipolar field. The different modes are easily separated by taking $\chi_1=0^o$ for $m=0$ whatever $(\chi_2,\chi_3,\chi_4)$, $(\chi_1,\chi_2)=(90^o,0^o)$ for $m=1$ whatever $\chi_3,\chi_4$, $(\chi_1,\chi_2,\chi_3)=(90^o,90^o,0^o)$ for $m=2$ whatever $\chi_4$, $(\chi_1,\chi_2,\chi_3,\chi_4)=(90^o,90^o,90^o,0^o)$ for $m=3$ and $(\chi_1,\chi_2,\chi_3,\chi_4)=(90^o,90^o,90^o,90^o)$ for $m=4$.

More handy expressions for these multipoles are given in the appendix~\ref{app:B} where  approximate solutions in the limit of a point multipole are shown (in the limit $R\to0$).
The retarded point multipoles represent very accurate and simple analytical expressions to study the effects of non dipolar fields on the structure of pulsar magnetospheres (or other non compact objects), its polar caps and its emission geometry. The detailed consequences of such fields, for instance on the radio polarization properties, requires a thorough analysis which we leave for future work. However, to get a flavour of multipolar effects, for the remainder of the paper, we will show the influence on the Poynting flux and related braking index as well as on the polar cap geometry and high energy emission phase diagrams.

\section{Poynting flux and braking index}
\label{sec:Poynting}

As a diagnostic of the consequences of the presence of multipole fields, we compute the exact Poynting flux of each multipole labelled by the mode $(l,m)$ taking into account the finite size of the star. The braking index~$n$ is an interesting related quantity which describes the efficiency of electromagnetic radiation by the law summarized in $\dot{\Omega} = - K \, \Omega^n$ where $K$ is a constant depending on boundary conditions on the neutron star. For magnetic multipolar point sources of order~$l$, we know that $n=2\,l+1$ \citep{1991ApJ...373L..69K} but in general it can differ from this fiducial value if the size of the star is taken into account.
Formally, the braking index follows from the luminosity~$L$ (or the power radiated) by the relation
\begin{equation}
 n = \frac{\Omega}{L} \, \frac{dL}{d\Omega} - 1
\end{equation}
which is derived from the definition of the braking index as
\begin{equation}
 n = \frac{\Omega\,\ddot{\Omega}}{\dot{\Omega}^2}
\end{equation}
and from the spin-down luminosity given by
\begin{equation}
 L = - I \, \Omega \, \dot{\Omega}
\end{equation}
where $I$ represents the moment of inertia of the neutron star. Note the minus sign in order to have $L>0$ because braking implies $\dot{\Omega}<0$.
Expressed in terms of the dimensionless parameter~$a=R/\rlight$ we get
\begin{equation}
\label{eq:IndiceFreinage}
 n = \frac{a}{L} \, \frac{dL}{da} - 1 = \frac{d\ln L}{d\ln a} - 1.
\end{equation}
This is the general formula to compute the braking index in any case, knowing the luminosity of the star with respect to the spin normalized by the parameter~$a$. Useful approximate expressions for the radiated power~$L$ and the braking index~$n$ are given in the following paragraphs starting from the dipole field and up to the octopole.

\subsection{Dipole}

\begin{figure*}
 \input{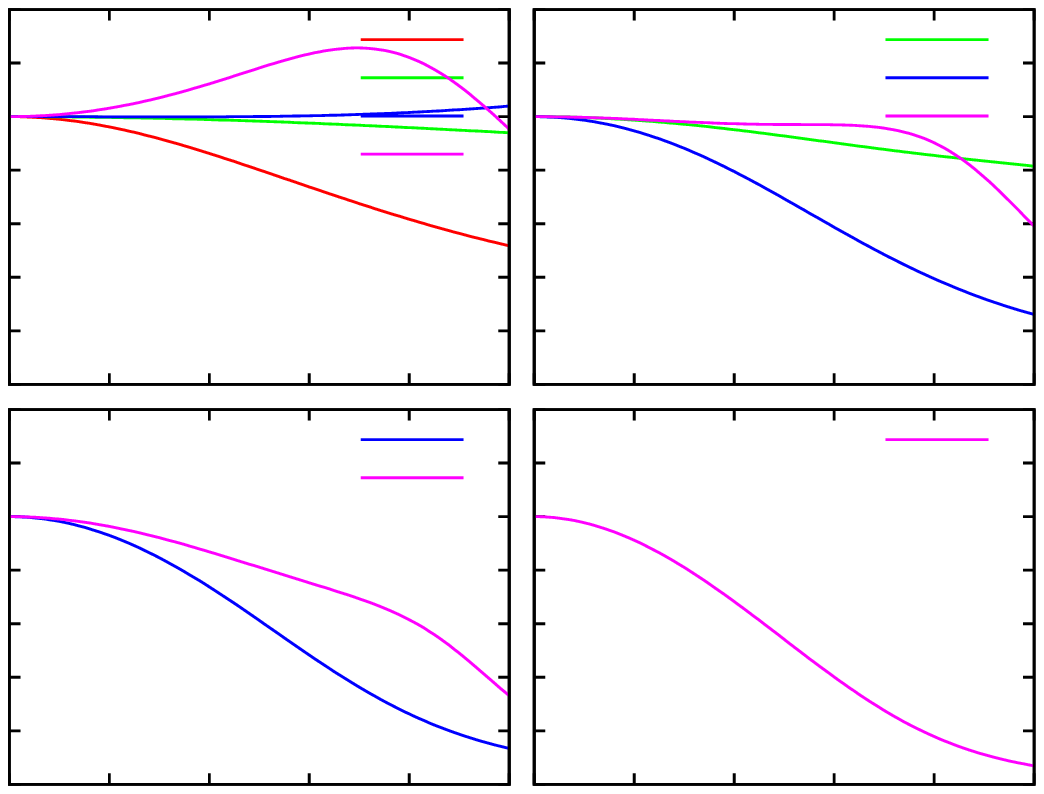}
 \caption{Spin-down luminosity of the rotating magnetic multipole according to the ratio $R/\rlight$. The contributions are separated into single components $(l,m)$. The dipole is shown in red, the quadrupole in green, the hexapole in blue and the octopole in magenta. Note that each luminosity is normalized to its value for $R=0$.}
 \label{fig:LuminositeMultipole}
\end{figure*}

\begin{figure*}
 \input{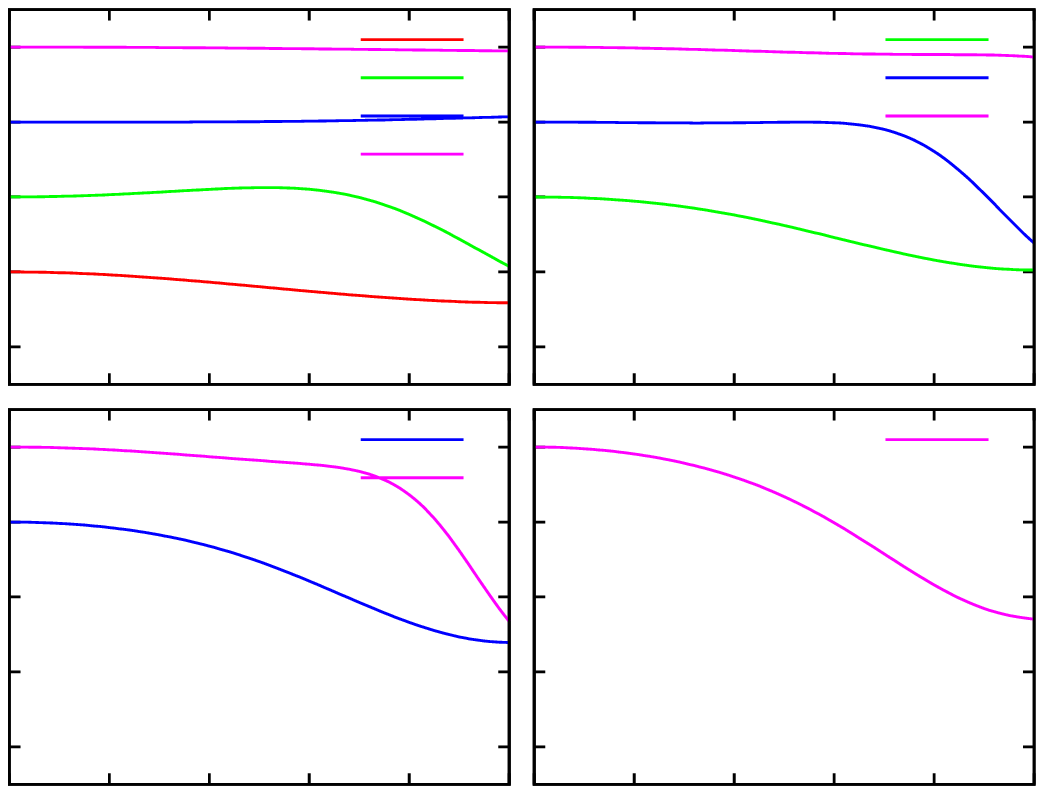}
 \caption{Braking index of the rotating magnetic multipole according to the ratio $R/\rlight$. The contributions are separated into single components $(l,m)$. The dipole is shown in red, the quadrupole in green, the hexapole in blue and the octopole in magenta.}
 \label{fig:IndiceFreinageMultipole}
\end{figure*}

First, we recall the exact Poynting flux obtained from Deutsch solution and compare it to the point dipole approximation. The magnetodipole point source losses are
\begin{equation}
 L^{\rm point}_{\rm dip} = \frac{8\,\upi}{3\,\mu_0\,c^3} \, \Omega^4\,B^2\,R^6 \, \sin^2\chi
\end{equation}
which can differ significantly from the true dipole losses generated by a star of finite size, according to the formula
\begin{subequations}
 \begin{align}
  \label{eq:LuminositeDipole}
 \frac{L_{\rm dip}}{L^{\rm point}_{\rm dip}} & = \frac{1}{a^2+1}+\frac{3 a^4}{5 \left(a^6-3 a^4+36\right)} \\
  \label{eq:LuminositeDipoleApprox}
 & \approx 1 - a^2 + o(a^2)
 \end{align}
\end{subequations}
Equation~(\ref{eq:LuminositeDipole}) is consistent with the expression given by~\cite{1997MNRAS.288.1049M}. From our general treatment of the multipole fields, we would write it as
\begin{equation}
 L_{\rm dip} = \frac{8\,\upi}{3\,\mu_0} \, c \, B^2\,R^2 \, \sin^2\chi \, \left[ \frac{1}{|h_1^{(1)}(k\,R)|^2} + \frac{3}{5} \, \frac{R^2}{\rlight^2} \, \frac{1}{|\partial_r ( r \, h_{2}^{(1)}(k\,r))|^2_R} \right]
\end{equation}
which is exactly the same as equation~(\ref{eq:LuminositeDipole}). The spin-down luminosity is always less than the point dipole source, its variation with respect to the stellar rotation speed is shown in Fig.~\ref{fig:LuminositeMultipole}, assuming that only the mode~$m=1$ is present ($\chi=90^o$). Because for all known pulsars, even for millisecond ones, we have $a\lesssim0.1$, the approximation to second order in $a$, as given by equation~(\ref{eq:LuminositeDipoleApprox}), is always sufficient to compute accurately the spin-down rate. The luminosity decreases monotonically with~$a$ reaching only 52\% of the point dipole source for $a=1$ although this rotation rate becomes unrealistic. Nevertheless it shows that the spin-down luminosity slightly changes with respect to~$\Omega$ and thus, rigorously, the braking index will deviate from the often quote value of three.
Indeed, this braking index derived from equation~(\ref{eq:IndiceFreinage}) is given by
\begin{subequations}
 \begin{align}
  \label{eq:IndiceDipole}
 n & = 1 + \frac{2}{a^2+1}+\frac{6 \left(a^4-45\right)}{2 a^6-3 a^4+45}-\frac{6 \left(a^4-36\right)}{a^6-3 a^4+36} \\
  \label{eq:IndiceDipoleApprox}
 & \approx 3 - 2 \, a^2 + o(a^2)
 \end{align}
\end{subequations}
The braking indexes given in this paper have been computed with the formal calculator \textit{Mathematica}. It is straightforward to check that eq.~(\ref{eq:IndiceDipole}) is exactly identical to eq.~(B4) of \cite{1997MNRAS.288.1049M} in the special case of $p=1$. Indeed, in our picture, the empty space corresponds to the region outside the star therefore using his parameter defined in his equation~(6) we have $x_\nu = R\,\Omega/c$ thus $\dot{x}_\nu/x_\nu = \dot{\Omega}/\Omega$ implying $p=1$ according to his definition.
The braking index strictly equals three only for the retarded point dipole. It is always less than three although not significantly except for fast rotating star for which $a\lesssim1$. For instance, for millisecond pulsars with $a=0.1$, the braking index is still as high as $n=2.98$. Its variations are shown in Fig.~\ref{fig:IndiceFreinageMultipole}. Starting from $n=3$ for $a=0$ it decreases to $n=2.18$ for $a=1$.

We must conclude that the Poynting flux and its braking index are subject to variations depending on the location of the inner boundary with respect to the light-cylinder as already mentioned by \cite{1997MNRAS.288.1049M}. This tells us that a braking index less than three can be explained by artificially increasing the size of the neutron star to such a point that it reaches almost its light-cylinder. This could happen for instance if the magnetosphere is only partially filled with electron/positron pairs, mostly in corotation with the star in a kind of dome+torus shape but also with an outer disk in overrotation. Such plasma equilibria are known for a long time and referred as electrospheres by \cite{1985MNRAS.213P..43K} and \cite{2002A&A...384..414P}. This argument is similar to the corotating inner magnetosphere of \cite{1997MNRAS.288.1049M} expect that in the latter model, the boundary is kept spherical whereas in the former, the boundary surface is much more complicated. The switch between a vacuum magnetosphere and an electrosphere leads to a switch in the Poynting flux and therefore could explain the switch-on/switch-off modes of several intermittent pulsars \citep{2006Sci...312..549K} although a quantitative analysis would require an accurate value of the spin-down power of the electrosphere, which unfortunately is not yet accessible.

\subsection{Quadrupole}

Let us investigate the Poynting flux emanating from a rotating quadrupolar magnetic field. The close analytical expression is given in terms of spherical Hankel functions such that
\begin{subequations}
\begin{align}
 L_{\rm quad} & = \frac{4\,\upi}{3\,\mu_0} \, c \, B^2\,R^2 \, \sin^2\chi_1 \, \\
 & \left[ \cos^2\chi_2 \, \left( \frac{1}{|h_2^{(1)}(k\,R)|^2} + \frac{R^2}{\rlight^2} \, \left( \frac{64}{35} \, \frac{1}{|\partial_r ( r \, h_{3}^{(1)}(k\,r))|^2_R} + \frac{3}{5} \, \frac{1}{|\partial_r ( r \, h_{1}^{(1)}(k\,r))|^2_R} \right) \right) + \right. \nonumber \\
 & \left. \sin^2\chi_2 \, \left( \frac{1}{|h_2^{(1)}(2\,k\,R)|^2} + \frac{R^2}{\rlight^2} \, \left( \frac{8}{7} \, \frac{1}{|\partial_r ( r \, h_{4}^{(1)}(2\,k\,r))|^2_R} \right) \right)\right] \nonumber
 \end{align}
\end{subequations}
In order to get more tractable expressions, we compute the point quadrupole losses given by
\begin{equation}
  L^{\rm point}_{\rm quad} = \frac{128\,\upi}{135\,\mu_0\,c^5} \, \Omega^6 \, B^2 \, R^8 \, \sin^2\chi_1 \, ( \cos^2\chi_2 + 10 \, \sin^2\chi_2 )
\end{equation}
It depends on $\sin^2\chi_1$, which is reminiscent of the dipole field, but now it also depends on the second angle~$\chi_2$ through a more complex trigonometric variation. The difference in spin-down luminosity between the configuration $\chi_2=0^o$ and $\chi_2=90^o$ reaches a factor ten. The exact explicit expression for the true quadrupolar losses being to lengthy, we give the second order approximation in $a$ which should be enough in any cases for pulsars. Compared to the point quadrupole, it is given by
\begin{equation}
  \frac{L_{\rm quad}}{L^{\rm point}_{\rm quad}} \approx 1+\frac{a^2 \left(301-339 \cos 2 \chi _2 \right)}{24 \left(9 \cos 2 \chi _2-11\right)} + o(a^2)
\end{equation}
However, the exact analytical variations of $L_{\rm quad}$ are depicted in Fig.~\ref{fig:LuminositeMultipole}. We separated the contributions from the mode~$m=1$ ($\chi_2=0^o)$ and the mode~$m=2$ ($\chi_2=90^o)$ and assume $\chi_1=90^o$. Both contributions decrease monotonically with $a$. Note that each luminosity is normalized to its value for $a=0$. The largest fluctuations in spin-down are less than 20\%.

To the same order of accuracy in $a$, the braking index is
\begin{equation}
 n \approx 5 + \frac{a^2}{12} \, \frac{19 \, \cos^2 \chi _2 - 320 \, \sin^2\chi_2}{\cos^2\chi_2 + 10 \, \sin^2\chi_2} + o(a^2)
\end{equation}
It does not depend on $\chi_1$ as it did not depend on $\chi$ for the dipole. It is always less than five for the mode $m=2$, decreasing monotonically as shown in Fig.~\ref{fig:IndiceFreinageMultipole}. On the contrary, the braking index for the mode $m=1$ increases slightly up to $n=5.25$ for $a\approx0.5$ before decreasing sharply down to $n=3.17$ for $a=1$.

\subsection{Hexapole}

The Poynting flux emanating from a rotating hexapolar magnetic field is given by a close analytical expression in terms of spherical Hankel functions, see equation~(\ref{eq:luminosite_un_multipole}). We do not write down its explicit expression because it is too lengthy.
Nevertheless, the point hexapole losses can be cast in short form as
\begin{subequations}
\begin{align}
 L^{\rm point}_{\rm hexa} & = \frac{8}{4725\,\mu_0\,c^7} \, \Omega^8\,B^2\,R^{10} \, \sin^2\chi_1 \\
 & ( 29 \, \cos^2\chi_2 + \sin^2\chi_2 \, ( 1664 \, \cos^2\chi_3 + 15309\, \sin^2\chi_3 ) ) \nonumber
\end{align}
\end{subequations}
It depends on $\sin^2\chi_1$, but also on the second and third angles~$(\chi_2,\chi_3)$ through a more complex trigonometric variation. The difference in spin-down luminosity between the configuration $(\chi_2,\chi_3)=\{(0^o,0^o),(90^o,0^o),(90^o,90^o)\}$ is in the ratio $29:1664:15309$. These configurations are respectively associated to the mode $m=1:2:3$.

The true hexapolar spin-down is approximately given to second order in $a$ by
\begin{equation}
  \frac{L_{\rm hexa}}{L^{\rm point}_{\rm hexa}} \approx 1-\frac{7 a^2 \left(2 \cos ^2\chi _2+\sin ^2\chi _2 \left(60073-58025 \cos \left(2 \chi _3\right)\right)\right)}{15 \left(58 \cos
   ^2\chi _2+\sin ^2\chi _2 \left(16973-13645 \cos \left(2 \chi _3\right)\right)\right)} + o(a^2)
\end{equation}
The exact variations are shown in Fig.~\ref{fig:LuminositeMultipole}. For the mode $m=1$, the spin-down slowly increases with $a$ up to $1.04$ whereas the modes $m=2$ and $m=3$ decrease with $a$ down to respectively $0.26$ and $0.13$.

The corresponding braking index is
\begin{subequations}
 \begin{align}
 n & \approx 7 - \frac{14\,a^2}{15} \, \frac{\cos^2 \chi _2 + \sin^2\chi_2 \, ( 1024 \, \cos^2\chi_3 + 59049 \, \sin^2\chi_3  )}{29 \, \cos^2\chi_2 + \sin^2\chi_2 \, ( 1664 \, \cos^2\chi_3 + 15309\, \sin^2\chi_3 )} + o(a^2)
 \end{align}
\end{subequations}
As shown in Fig.~\ref{fig:IndiceFreinageMultipole}, its is less than seven for the modes $m=2$ and $m=3$, monotonically decreasing to $n=3.77$ and $n=3.78$ respectively. For the mode $m=1$, the braking index is almost constant, $n\approx7$, showing a slight increase up to $n=7.13$.

\subsection{Octopole}

Eventually, the Poynting flux from a rotating octopolar magnetic field is studied.
The point octopole losses are
\begin{subequations}
\begin{align}
 L^{\rm point}_{\rm octo} & = \frac{16}{297675\,\mu_0\,c^9} \, \Omega^{10}\,B^2\,R^{12} \, \sin^2 \chi_1 \, \\
 & ( 23 \, \cos^2\chi_2 + \sin^2\chi_2 \, ( 5632 \, \cos^2\chi_3 + \sin^2\chi_3 \, ( 133407 \, \cos^2\chi_4 + 11179648 \, \sin^2\chi_4 ) ) )
\end{align}
\end{subequations}
The dependence on $\sin^2\chi_1$ is also present, and we retrieve also the second, third and fourth angles~$(\chi_2,\chi_3,\chi_4)$ describing the octopole configuration. The difference in spin-down luminosity between the configuration $(\chi_2,\chi_3,\chi_4)=\{(0^o,0^o,0^o),(90^o,0^o,0^o),(90^o,90^o,0^o),(90^o,90^o,90^o)\}$ is in the ratio $23:5632:133407:11179648$ (which are for the mode $m=1:2:3:4$ respectively). The mode $m=4$ has the maximal spin-down rate, several orders of magnitude larger than the mode $m=1$. Indeed, in all multipole fields, the largest spin-down luminosity is attained for $l=m$.

The true octopolar spin-down losses to second order in $a$ is given by
\begin{equation}
  \frac{L_{\rm octo}}{L^{\rm point}_{\rm oct}} \approx 1-\frac{a^2 \left(68 \cos ^2\chi _2+2 \sin ^2\chi _2 \left(38912 \cos ^2\chi _3+27 \sin ^2\chi _3
   \left(1097419-999733 \cos \left(2 \chi_4\right)\right)\right)\right)}{21 \left(46 \cos ^2\chi _2+\sin ^2\chi _2 \left(11264
   \cos ^2\chi _3+9 \sin ^2\chi _3 \left(145895-116249 \cos \left(2 \chi _4\right)\right)\right)\right)} + o(a^2)
\end{equation}
The exact variations are shown in Fig.~\ref{fig:LuminositeMultipole}. For the mode $m=1$, the spin-down significantly increases with $a$ up to $1.24$ for $a\approx0.7$ and then decreases down to $0.95$.  All other modes $m=2$, $m=3$ and $m=4$ decrease with $a$ down to respectively $0.59$, $0.33$ and $0.07$.
This is a general trend.

The corresponding braking index is
\begin{subequations}
 \begin{align}
 n & \approx 9 - \frac{4\,a^2}{21} \, \frac{17 \, \cos^2 \chi _2 + \sin^2\chi_2 \, ( 19456 \, \cos^2\chi_3 + \sin^2\chi_3 \, ( 1318761 \cos^2\chi_4 + 28311552 \, \sin^2\chi_4 ) )}{23 \, \cos^2\chi_2 + \sin^2\chi_2 \, ( 5632 \, \cos^2\chi_3 + \sin^2\chi_3 \, ( 133407 \, \cos^2\chi_4 + 11179648 \, \sin^2\chi_4 ) )} + o(a^2)
 \end{align}
\end{subequations}
Its exact variations are shown in Fig.~\ref{fig:IndiceFreinageMultipole}. The braking index always decreases, whatever the mode $m$. The final value for $a=1$ can be significantly less than nine. Indeed, for $m=4$ the lowest value is $n=4.41$ and for $m=3$ it is $n=4.34$. 

In this paragraph, we demonstrated that the finite size of the magnetic multipole can drastically alter the point multipole picture. This reflects in the Poynting flux and therefore also in the derived braking index showing sometimes strong decrease with increasing rotation rate. These effects are rooted in the boundary conditions on the stellar surface. We must conclude that the electrodynamics on the neutron star surface affects the long term evolution of its rotation. As a matter of fact, on non spherical boundary such as those obtained from the electrosphere, results would deviate from the pure point source significantly. Explaining the behaviour of intermittent pulsars certainly requires a better and more quantitative treatment of those boundary surfaces which should be included with caution.

\subsection{Implications for the magnetic field strength estimates}

Taking into account the above spin-down luminosities for the multipole fields, we conclude that the magneto-dipole losses often used to estimate the magnetic field strength at the neutron star surface is irrelevant if small scale structures exist in the crust like the ones induced by high multipole components. The dipole formula only gives a good estimate of the magnetic field strength in the far zone, far outside the light-cylinder, but will not furnish a reliable estimate close to the neutron star surface where multipolar components are suspected to be dominant. 

The luminosities of the multipoles separated into their components of order $(l,m)$ are summarized in Table~\ref{tab:Spindown}. The luminosity is normalized according to the mode $m=1$. The dependence on the physical parameters $(B,\Omega,R)$ is given in the last column.
\begin{table}
 \centering
 \begin{center}
\begin{tabular}{cccccc}
 \hline
 $l/m$ & 1 & 2 & 3 & 4 & Normalization \\
 \hline
 \hline
 1 & $1$ &  &  & & $\displaystyle \frac{8\,\upi}{3} \, \frac{B^2\,\Omega^4\,R^6}{\mu_0\,c^3}$ \\
 2 & $1$ & $10$ &  & & $\displaystyle \frac{128\,\upi}{135} \, \frac{B^2\,\Omega^6\,R^8}{\mu_0\,c^5}$ \\
 3 & $1$ & $\displaystyle \frac{1664}{29}$ & $\displaystyle \frac{15309}{29}$ & & $\displaystyle \frac{232\,\upi}{4725} \, \frac{B^2\,\Omega^8\,R^{10}}{\mu_0\,c^7}$ \\
 4 & $1$ & $\displaystyle \frac{5632}{23}$ & $\displaystyle \frac{133407}{23}$ & $\displaystyle \frac{1179648}{23}$ & $\displaystyle \frac{368\,\upi}{297675} \, \frac{B^2\,\Omega^{10}\,R^{12}}{\mu_0\,c^{9}}$ \\
 \hline
 \end{tabular}
 \end{center}
 \caption{Spin-down luminosity for the point multipole source $\mathcal{L}(l,m)$, normalized to its value for $m=1$.}
 \label{tab:Spindown}
\end{table}
In order to get an idea of the misleading field strengths obtained by including only the dipole, we estimate the magnetic field on the surface by equating the multipole spin down of order $(l,m)$ to the rotational luminosity given by $\dot E = I \,\Omega\,\dot\Omega$ where $I$ is the moment of inertia of the star. A typical value used here is $I=10^{38}\textrm{ kg m}^2$. The fiducial pulsar parameters used are $P=1$~s and $\dot P=10^{-15}$. Results are summarized in Table~\ref{tab:MagneticStrength}.
\begin{table}
 \centering
 \begin{center}
\begin{tabular}{cccccc}
 \hline
 $l/m$ & 1 & 2 & 3 & 4 & Normalization \\
 \hline
 \hline
 1 & $1\times10^{8}$ &  &  & & $P^{1/2} \, \dot P_{-15}^{1/2}$ \\
 2 & $8\times10^{11}$ & $3\times10^{11}$ &  & & $P^{3/2} \, \dot P_{-15}^{1/2}$ \\
 3 & $2\times10^{16}$ & $2\times10^{15}$ & $7\times10^{14}$ & & $P^{5/2} \, \dot P_{-15}^{1/2}$ \\
 4 & $5\times10^{20}$ & $3\times10^{19}$ & $7\times10^{18}$ & $2\times10^{18}$ & $P^{7/2} \, \dot P_{-15}^{1/2}$ \\
 \hline
 \end{tabular}
 \end{center}
 \caption{Magnetic multipole strength in Tesla according to the pulsar spin-down luminosity. The fiducial parameters are $P=1$~s and $\dot P=10^{-15}$.}
 \label{tab:MagneticStrength}
\end{table}
The dipole is the most efficient radiator in the sense that it needs the lowest magnetic field to achieve a certain amount of spin-down luminosity. The next most efficient radiator is the quadrupole with $l=m=2$. For a given multipole of order $l$, the sectorial mode $l=m$ is the most powerful radiator. While a dipole needs a magnetic field of the order $B\approx10^8$~T to account for a given spin-down, a quadrupole needs $B\approx10^{11}$~T, a hexapole $B\approx10^{14}$~T and a octopole $B\approx10^{18}$~T. High multipoles can have magnetic fields above the critical field of $B_q=4.4\times10^9$~T without contradicting the requirement of dominant dipole losses.
Even if multipole spin-down can be negligible compared to the dipole, its multipolar magnetic field can be much larger than the dipole field. The upper limits for multipole fields $B_{l,m}$ seems not to be very restrictive for normal pulsars.

For millisecond pulsars the situation is different. Having typical period and period derivative parameters of $P=10^{-3}$~s and $\dot P=10^{-18}$ respectively, the constraints on multipoles are tighter. Results are summarized in Table~\ref{tab:MagneticStrengthMillisecond}. The magnetic field is always less than the critical field $B_q$, at least for $l\leqslant4$.
\begin{table}
 \centering
 \begin{center}
\begin{tabular}{cccccc}
 \hline
 $l/m$ & 1 & 2 & 3 & 4 & Normalization \\
 \hline
 \hline
 1 & $1\times10^{5}$ &  &  & & $P_{-3}^{1/2} \, \dot P_{-18}^{1/2}$ \\
 2 & $8\times10^{5}$ & $3\times10^{5}$ &  & & $P_{-3}^{3/2} \, \dot P_{-18}^{1/2}$ \\
 3 & $2\times10^{7}$ & $2\times10^{6}$ & $7\times10^{5}$ & & $P_{-3}^{5/2} \, \dot P_{-18}^{1/2}$ \\
 4 & $5\times10^{8}$ & $3\times10^{7}$ & $7\times10^{6}$ & $2\times10^{6}$ & $P_{-3}^{7/2} \, \dot P_{-18}^{1/2}$ \\
 \hline
 \end{tabular}
 \end{center}
 \caption{Magnetic multipole strength in Tesla according to the pulsar spin-down luminosity. The fiducial parameters are those for millisecond pulsars with $P=10^{-3}$~s and $\dot P=10^{-18}$.}
 \label{tab:MagneticStrengthMillisecond}
\end{table}

The magnetodipole losses formula should be avoided to guess the magnetic field strength at the surface of the neutron star. The magnetic field coming from the multipole components are completely missed, although they can be as large or even larger than the dipole one, nevertheless without contributing much to the total spin-down power. Observations of the pulsar spin-down alone is not reliable to estimate the magnetic field strength at its surface.

We go on with a discussion about the geometrical effects of multipole fields. Because it becomes very cumbersome to investigate in detail the influence of each multipole, having an increasing number of free parameters with increasing $l$, we restrict ourself to the dipole plus quadrupole moments.

\begin{figure}
 \centering
 \input{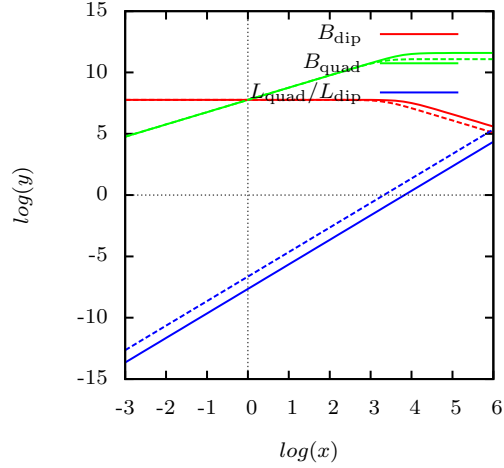}
 \caption{Ratio of quadrupole to dipole Poynting flux (blue line), intensity of dipolar field (red line), intensity of quadrupolar field (green line) with respect to $x$ for pulsars with a period of one second. Solid lines correspond to $X=1$ (m=1) and dashed lines to $X=10$ (m=2). Note the log scales on both axis and that magnetic field strengths are given in Tesla. The black dotted vertical line depicts the equality in field intensities $B_{\rm dip} = B_{\rm quad}$. The black dotted horizontal line depicts the equality in Poynting fluxes $L_{\rm dip} = L_{\rm quad}$.}
 \label{fig:DipvsQuadSecond}
\end{figure}

\begin{figure}
 \centering
 \input{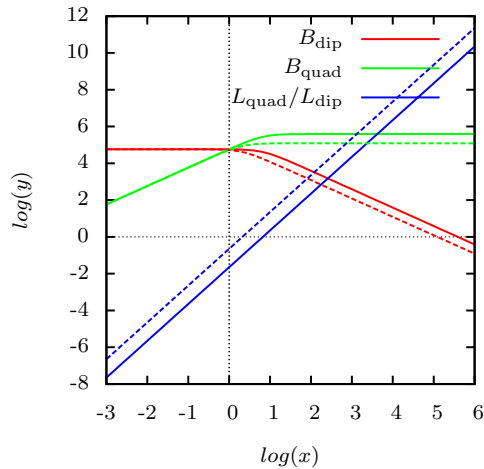}
 \caption{Ratio of quadrupole to dipole Poynting flux (blue line), intensity of dipolar field (red line), intensity of quadrupolar field (green line) with respect to $x$ for pulsars for millisecond pulsars. Solid lines correspond to $X=1$ (m=1) and dashed lines to $X=10$ (m=2). Note the log scales on both axis and that magnetic field strengths are given in Tesla. The black dotted vertical line depicts the equality in field intensities $B_{\rm dip} = B_{\rm quad}$. The black  dotted horizontal line depicts the equality in Poynting fluxes $L_{\rm dip} = L_{\rm quad}$.}
 \label{fig:DipvsQuadMilliSecond}
\end{figure}

An illustrative example is given by a comparison between the magneto-dipole and magneto-quadrupole losses. We compare specifically both $m=1$ modes or the $m=1$ dipole with the $m=2$ quadrupole knowing from the above discussion that
\begin{subequations}
 \begin{align}
  L_{\rm dip} & = \frac{8\,\upi}{3} \, \frac{B_{\rm dip}^2 \,\Omega^4\,R^6}{\mu_0\,c^3} \\
  L_{\rm quad} & = \frac{128\,\upi}{135} \, X \, \frac{B_{\rm quad}^2 \,\Omega^6\,R^8}{\mu_0\,c^5}
 \end{align}
\end{subequations}
where $X=1$ for $m=1$ and $X=10$ for $m=2$, see table~\ref{tab:Spindown}.
Introducing the ratio between the magnetic quadrupole field~$B_{\rm quad}$ and the magnetic dipole field~$B_{\rm dip}$ by
\begin{equation}
 x = \frac{B_{\rm quad}}{B_{\rm dip}}
\end{equation}
the associated Poynting flux is a quadratic function of $x$ such that
\begin{equation}
 \frac{L_{\rm quad}}{L_{\rm dip}} = \frac{16\,X}{45} \, x^2 \, \frac{R^2}{\rlight^2}.
\end{equation}
Assuming that the relevant parameters $(P,\dot P)$ are known for each pulsar, we can deduce the dipolar part by fixing $x$ and get
\begin{equation}
 B_{\rm dip} = \sqrt{\frac{3\,\mu_0\,c^3\,\dot E}{8\,\upi\,\Omega^4\,R^6 \, (1 + \frac{16\,X}{45} \, x^2 \, \frac{R^2}{\rlight^2})}}
\end{equation}
where the spindown is $\dot E=4\,\upi^2\,I\,\dot P \, P^{-3}$ and $I$ the neutron star moment of inertia taken to be $I=10^{38} \textrm{ kg\,m}^2$. We show the variations of $B_{\rm dip}$, $B_{\rm quad}$ and $L_{\rm quad}/L_{\rm dip}$ for typical pulsar parameters corresponding to a one second period in Fig.~\ref{fig:DipvsQuadSecond} and to a millisecond period in Fig.~\ref{fig:DipvsQuadMilliSecond}.
Inspecting Fig.~\ref{fig:DipvsQuadSecond}, for normal pulsars, we recognize that a dominant dipolar Poynting flux does not imply a dominant dipolar magnetic field. Indeed for $x=1$ the quadrupolar flux is still six to seven orders of magnitude less than the dipole flux. They become comparable only for $x=10^3$. If we assume that the Poynting flux remains mainly dipolar, we are unable to put severe constraints on the magnitude of multipolar fields such as the quadrupolar components (m=1,2). The magneto-dipole losses are useless to get drastic upper limits for multipole components. This conclusion is even more severe for magnetars possessing periods longer than one second.

The situation changes for millisecond pulsars, Fig.~\ref{fig:DipvsQuadMilliSecond}. For $x=1$, dipolar and quadrupolar Poynting fluxes are comparable the latter becoming dominant above $x=1$. Thus if we assume that the dipole spindown remains dominant, we get stringent upper limits for at least the quadrupolar moment, being $B_{\rm quad} \lesssim B_{\rm dip}$.

\section{Polar cap geometry}
\label{sec:PolarCap}

It is believed that the geometry of the polar caps determines the radio pulse profiles as the coherent emission is assumed to be produced close to the neutron star surface, at heights comparable to the neutron star radius. This region is therefore privileged to look for possible significant multipolar components in the magnetic field because a multipole of order~$l$ behaves like $r^{-(l+2)}$ near the origin.

In order to characterize the quadrupole with respect to the dipole, we introduce two weights $w_{\rm d}$ and $w_{\rm q}$ such that the total magnetic field is equal to a weighted linear combination
\begin{equation}
\label{eq:Bdipquad}
 B = w_{\rm d} \, B_{\rm dip} + w_{\rm q} \, B_{\rm quad} .
\end{equation}
We use the retarded point multipole expressions given in the appendix~\ref{app:B}. Nevertheless, we assume that the neutron star radius is equal to $R/\rlight=0.1$ to compute the magnetic field lines.

\begin{figure}
 \input{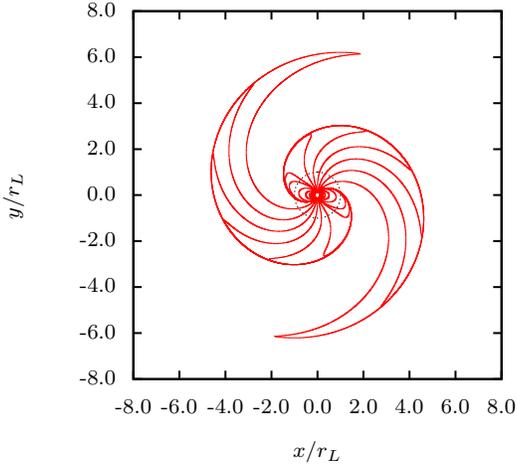}
\caption{Geometry of the magnetic field lines in the equatorial plane of an orthogonal point dipole. The dotted circle corresponds to the light-cylinder. The field lines approach only asymptotically the double Archimedes spirals (not shown).}
\label{fig:LigneDipole}
\end{figure}

\begin{figure}
 \input{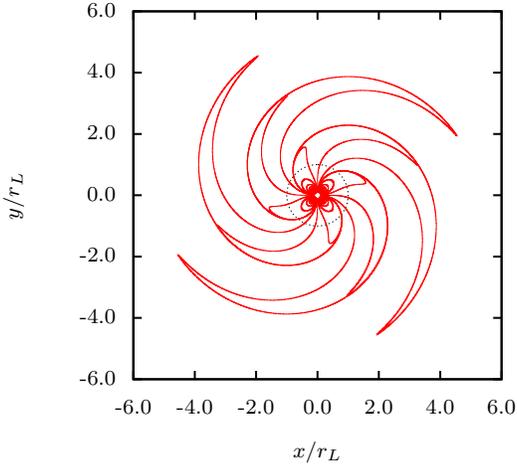}
\caption{Geometry of the magnetic field lines in the equatorial plane of an orthogonal point quadrupole. The dotted circle corresponds to the light-cylinder. The field lines approach only asymptotically the quadruple Archimedes spirals (not shown).}
\label{fig:LigneQuadrupole}
\end{figure}

To get an idea of the field line topology of the quadrupole compared to the dipole, we show some magnetic field lines contained in the equatorial plane for the orthogonal point dipole ($\chi=90^o$), fig.~\ref{fig:LigneDipole}, and the orthogonal point quadrupole ($(\chi_1,\chi_2)=(90^o,90^o)$), fig.~\ref{fig:LigneQuadrupole}. The wave like structure is easily recognized from respectively the double and the quadruple spiral pattern. Note however that the magnetic field lines approach only asymptotically these spiral patterns that are not shown in the figures for clarity.

In order to quantify how the quadrupolar field affects the number and geometry of the caps in the vicinity of the surface, we show several maps locating these polar caps. The physically relevant geometry depends on the relative weights between these dipolar and quadrupolar fields. 

\begin{figure*}
 \input{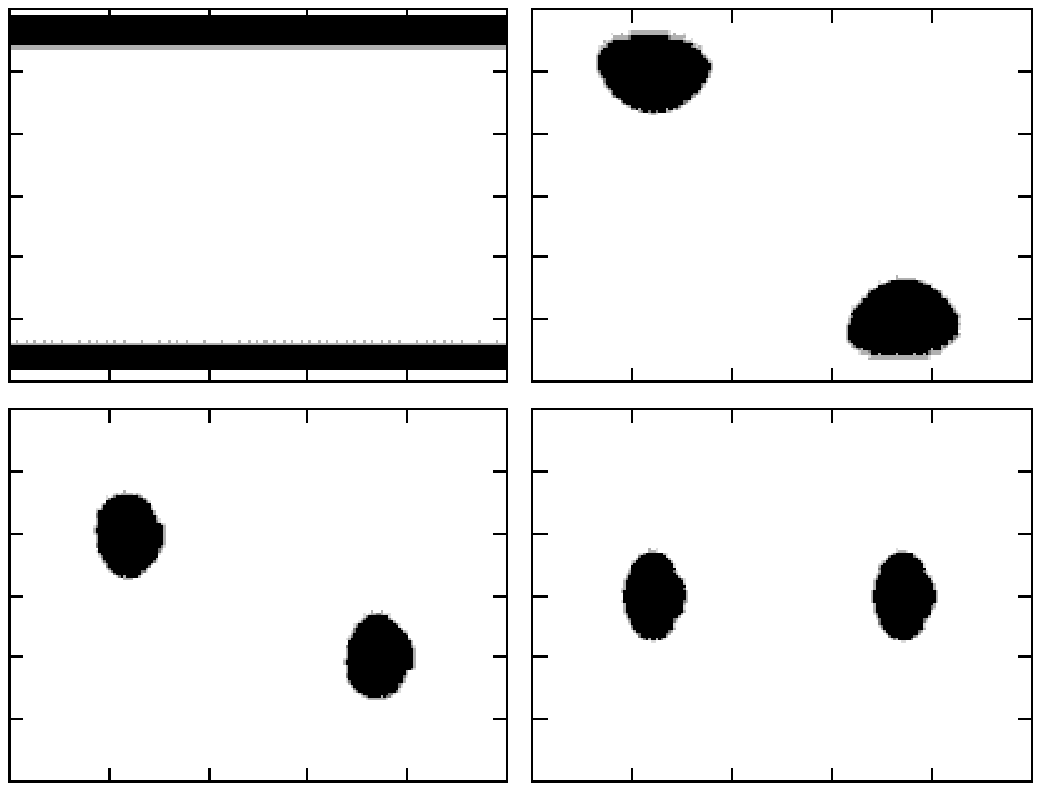}
\caption{Geometry of the polar cap for the retarded point dipole field solely with $R/\rlight=0.1$. The four panels correspond to different inclination angles~$\chi=\{0^o,30^o,60^o,90^o\}$ as indicated in the plots.}
\label{fig:PolarCapDipole}
\end{figure*}

To start with, we recall the polar caps for the dipole field as shown in Fig.~\ref{fig:PolarCapDipole} for four obliquities: $\chi=\{0^o, 30^o, 60^o,90^o\}$. The weights are $w_{\rm d}=1, w_{\rm q}=0$. The special aligned case with $\chi=0^o$ represents an azimuthally symmetric situation. Thus the polar cap location does not vary with the phase of the pulsar. In the misaligned case, two polar caps are always clearly seen. They are directly connected to both magnetic poles.

\begin{figure*}
 \input{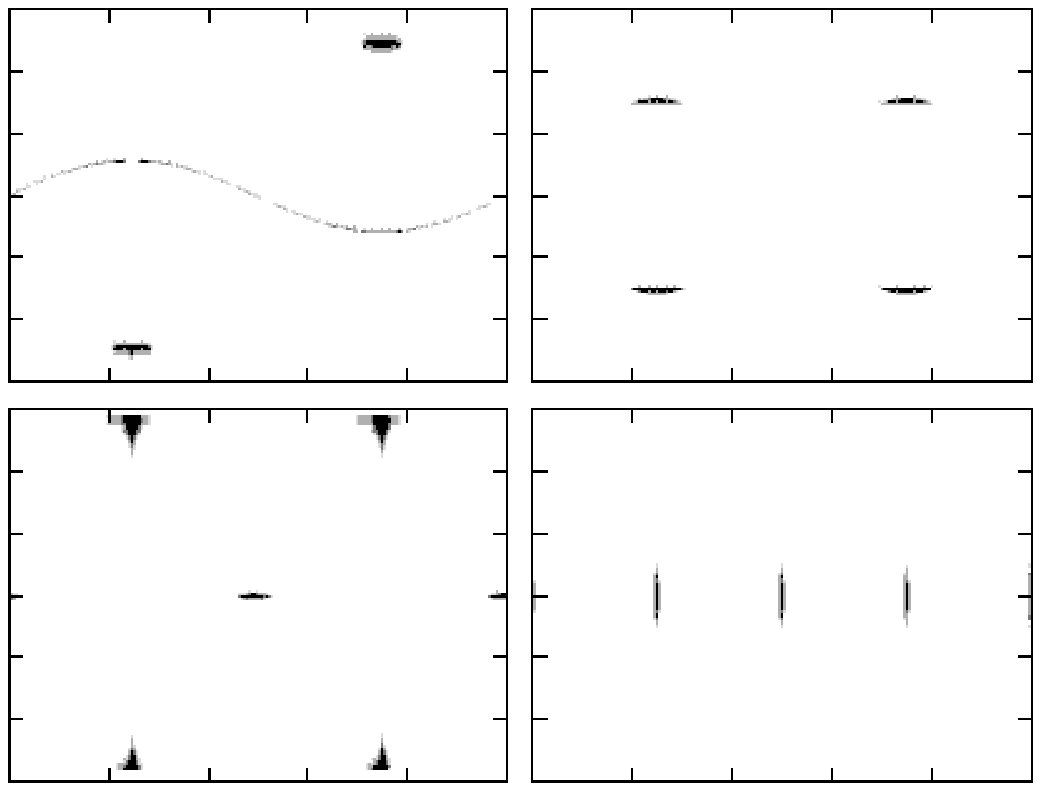}
\caption{Geometry of the polar cap for the quadrupole field solely with $R/\rlight=0.1$. The four panels correspond to different configurations depicted by the couple of angles~$(\chi_1, \chi_2)$ as indicated in the plots.}
\label{fig:PolarCapQuadrupole}
\end{figure*}

Next we consider a quadrupole alone. The weights are $w_{\rm d}=0, w_{\rm q}=1$. Because this field possesses four poles, we would expect at most four polar caps depending on the respective weight of each mode $m=\{0,1,2\}$. A typical sample of polar cap maps is shown in Fig.~\ref{fig:PolarCapQuadrupole}. The case $\chi_1=\chi_2=90^o$ is the most easiest to interpret. It is a mode $m=2$ with magnetic poles locate in the equatorial plane. Thus two couples of two poles with opposite polarity are present. These are easily identified in Fig.~\ref{fig:PolarCapQuadrupole}. Note however that one cap is hardly seen because of our phase origin which is exactly aligned with one pole. Thus one is centred at phase $\varphi=0$. If $\chi_1$ is decreased for instance to $\chi_1=30^o$, two of these poles move towards the rotation axis while the two other stay at the equator, see panel $(\chi_1=30^o,\chi_2=90^o)$. Next the case $(\chi_1=90^o,\chi_2=0^o)$ represents the mode $m=1$ alone. The four caps are still present, but they distribute symmetrically with respect to the equatorial plane. If $\chi_1$ is again decreasing, two caps move to the rotation axis whereas the two other move to the equatorial plane until they merge to form only one cap seen at any phase~$\varphi$, see panel $(\chi_1=30^o,\chi_2=0^o)$.

\begin{figure*}
 \input{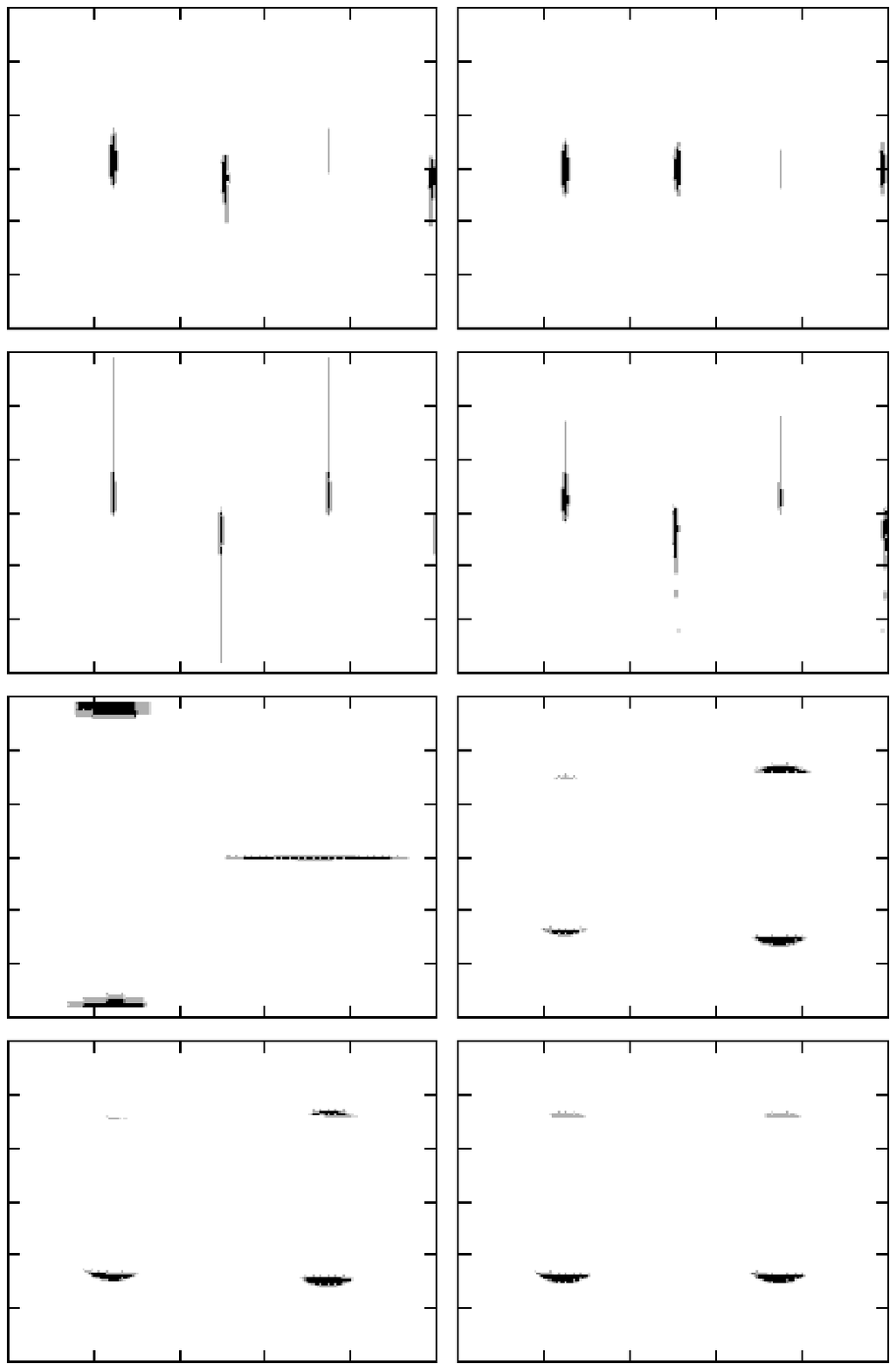}
\caption{Geometry of the polar cap for a linear combination of the dipole and quadrupole fields with $R/\rlight=0.1$ and weights $w_{\rm d} = w_{\rm q} = 1$. Each panel corresponds to a different configuration depicted by the angles~$(\chi, \chi_1, \chi_2)$. Their precise values are indicated in the plots.}
\label{fig:PolarCapDipoleQuadrupole}
\end{figure*}

Finally, we assume a linear combination of dipole and quadrupole fields as shown in Fig.~\ref{fig:PolarCapDipoleQuadrupole}. For concreteness, we choose weights such that $w_{\rm d}=1, w_{\rm q}=1$. Any other weights would also be acceptable but it is impossible to scan all relevant parameters and to show the results in a paper of a decent size. Certainly an atlas containing hundreds of maps could be computed but we do not show it here. We summarize our results by showing three different classes of behaviour. The first class starts with a quadrupole mode $m=2$ thus with $(\chi_1=90^o,\chi_2=90^o)$, see Fig.~\ref{fig:PolarCapQuadrupole}. We add to it a dipole with inclination going from an aligned rotator to a perpendicular rotator. We always find four polar caps located anti-symmetrically with respect to the equatorial plane. The configuration becomes exactly symmetric only for a perpendicular dipole, see $(\chi=90^o,\chi_1=90^o,\chi_2=90^o)$. In the other extreme case, see $(\chi=0^o,\chi_1=90^o,\chi_2=90^o)$, the polar caps are at the opposite of what would be expected for an aligned dipole alone. Indeed, their location contradicts the usual fact that an almost aligned rotator can only emit one pulse per period. With a quadrupole component, the prediction changes drastically. Two pulses per period, separated by half a period, would be detected and most importantly interpreted as a perpendicular (dipole) rotator. This is a far reaching important result of this paragraph. If multipole fields exist in neutron star, and we could not understand why they should not, all the geometrical fitting parameters obtained by interpreting the radio pulses would be inaccurate or worst, wrong.

The second case assumes a quadrupolar symmetric mode $m=0$ to which we add a perpendicular dipole as seen in the panel $(\chi=90^o,\chi_1=0^o,\chi_2=0^o)$. This configuration depicts the opposite example of the previous case, in which a clear misinterpretation of perpendicular rotator being an aligned rotator is performed. We always only see one pulse per period with a duty cycle between 20\% and 50\%. This behaviour is usually explained by an almost aligned rotator with $\chi$ close to zero. To the contrary, here, the dipole is perpendicular but the quadrupole is aligned.

The third and last case corresponds to the quadrupole mode $m=1$. Thus we add a $m=1$ dipole to a $m=1$ quadrupole. We always observe four polar caps. For too small obliquity $\chi\lesssim45^o$, on polar cap disappears, only three remain, see panel~$(\chi=0^o,\chi_1=90^o,\chi_2=0^o)$. For sufficiently large obliquity $\chi\gtrsim45^o$, polar caps located in the same hemisphere become identical.

Before dealing with the high-energy emission properties, we stress the main important result of this paragraph. If pulsar magnetospheres are only described with a misaligned dipolar magnetic field, interpretation of the radio pulse profile becomes extremely misleading for some special cases. Taking into account quadrupole or higher order multipole fields would destroy the fitting obtained by a dipole solely and the value of the parameters would be questionable.

\section{High-energy emission phase diagram}
\label{sec:Emission}

The multipolar magnetic field impacts the emission in the whole electromagnetic spectrum, not only radio as shown in the previous section but also the high energy counterpart. This is especially true for millisecond pulsars, those having a large $R/\rlight$ ratio. In order to show the complexity of light curves emanating from such magnetic configurations, we plot several phase diagrams including dipolar and quadrupolar fields possessing similar relative strengths ($w_{\rm d} = w_{\rm q} = 1$). In order to compute the light-curves we used a slot gap model as described in \cite{2015A&A...574A..51P}.

\begin{figure*}
 \input{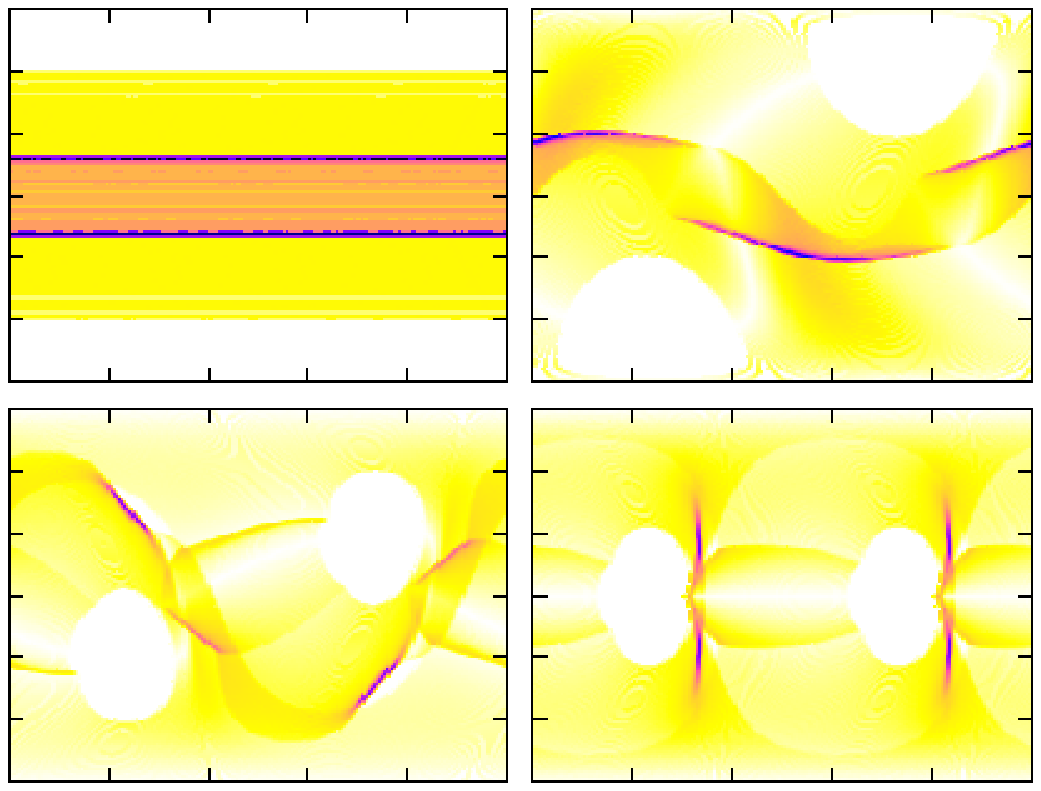}
\caption{Phase diagram for the retarded point dipole rotator with $R/\rlight=0.1$. The four panels correspond to different inclination angles~$\chi=\{0^o,30^o,60^o,90^o\}$ as indicated in the plots.}
\label{fig:PhaseDiagramDipole}
\end{figure*}

The phase diagram for the dipole is shown in Fig.~\ref{fig:PhaseDiagramDipole}. The caustic effect is seen in red-blue-black colour, corresponding to the highest intensity. The non emitting phases remain in white. They are larger than the related polar caps because the magnetic field lines are not directed radially but their opening angle is larger than the cone subtended by the polar cap by approximately 50\% \citep{2001ApJ...555...31G}. 

\begin{figure*}
 \input{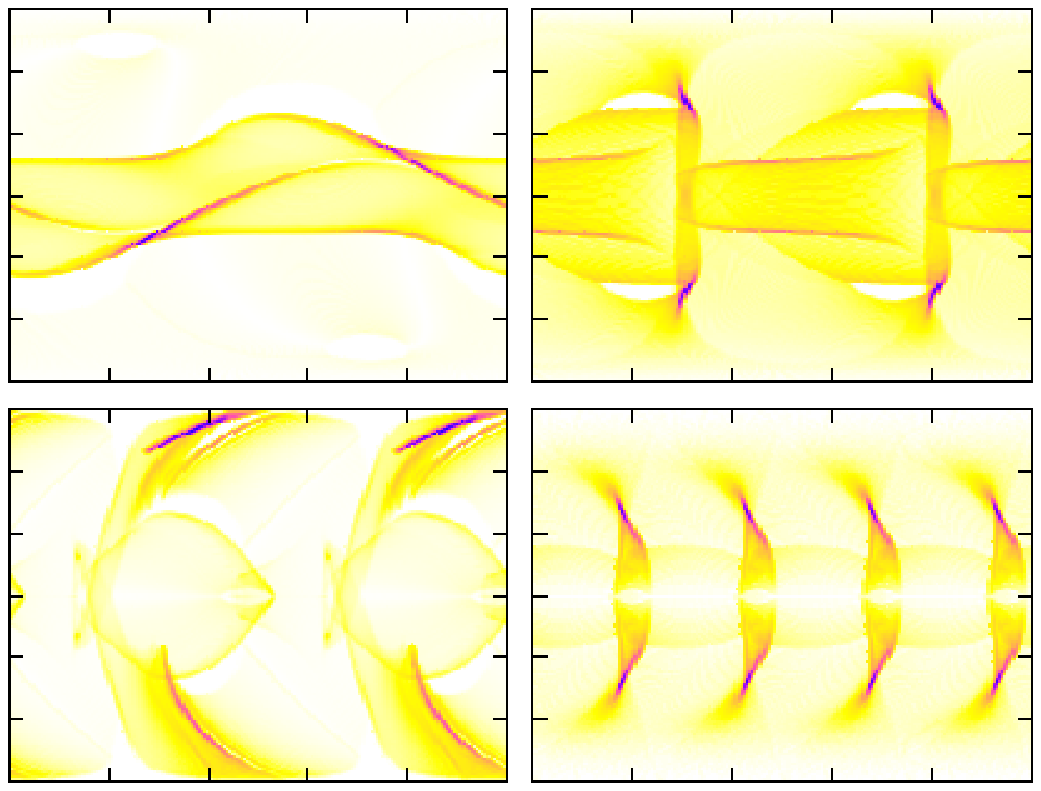}
\caption{Phase diagram for the retarded point quadrupole rotator with $R/\rlight=0.1$ and inclination angles $(\chi_1, \chi_2)$.}
\label{fig:PhaseDiagramQuadrupole}
\end{figure*}

For the quadrupole alone, we recognize an $m=1$ structure for $(\chi_1=90^o,\chi_2=0^o)$ with two main peaks whereas for a $m=2$ structure with $(\chi_1=90^o,\chi_2=90^o)$ we observe four peaks, see Fig.~\ref{fig:PhaseDiagramQuadrupole}. Two other intermediate cases are also shown corresponding to the same configuration as in Fig.~\ref{fig:PolarCapQuadrupole}. Note that in general the north-south symmetry is broken by the presence of a quadrupole. The case $(\chi_1=30^o,\chi_2=0^o)$ possesses a significant bridge emission between both pulses around $\zeta=90^o$.

\begin{figure*}
 \input{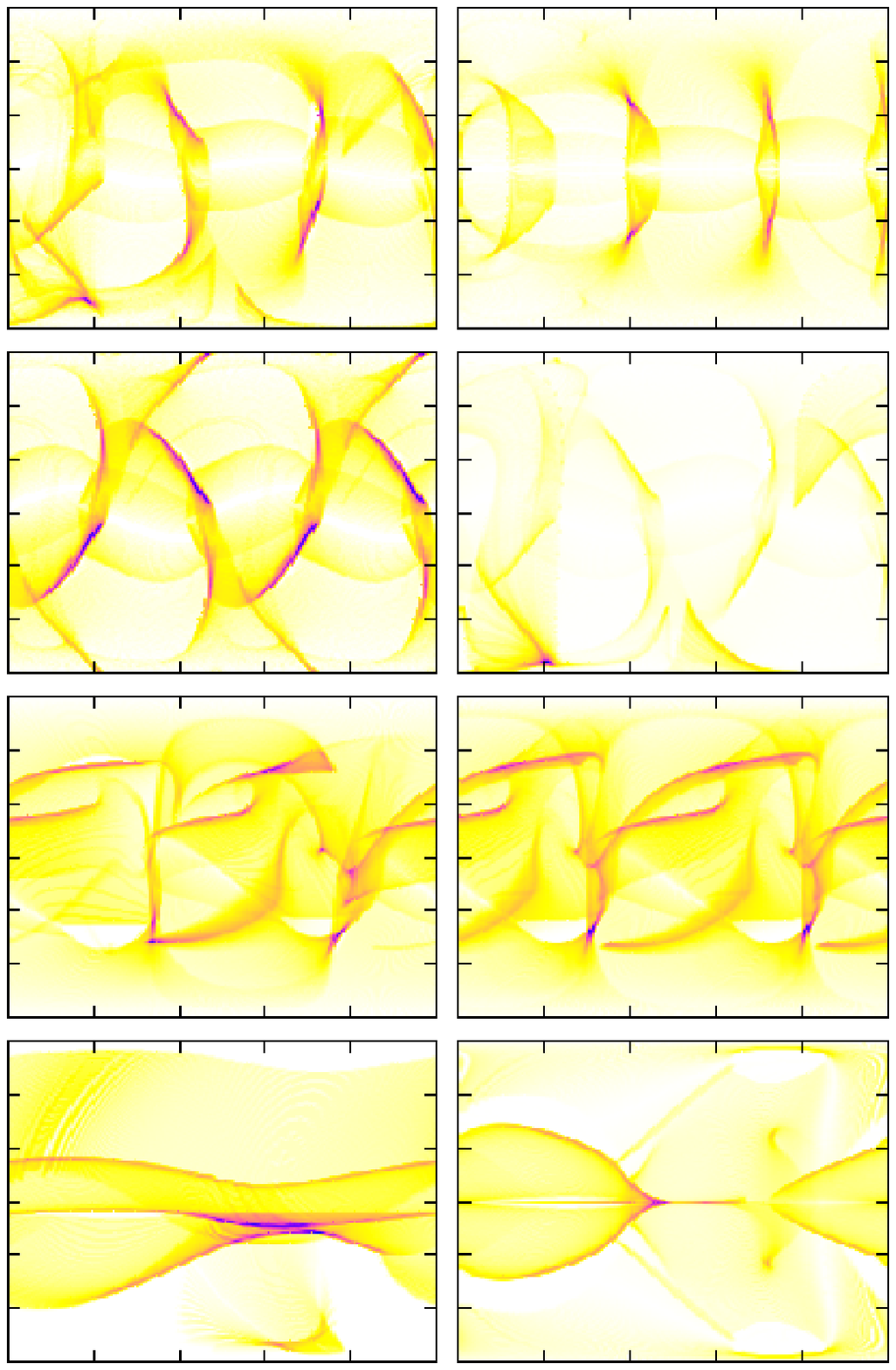}
\caption{Phase diagram for the retarded point dipole and quadrupole and the inclination angles $(\chi, \chi_1, \chi_2)$ for weights $w_{\rm d} = w_{\rm q} = 1$.}
\label{fig:PhaseDiagramDipoleQuadrupole2}
\end{figure*}

\begin{figure*}
 \input{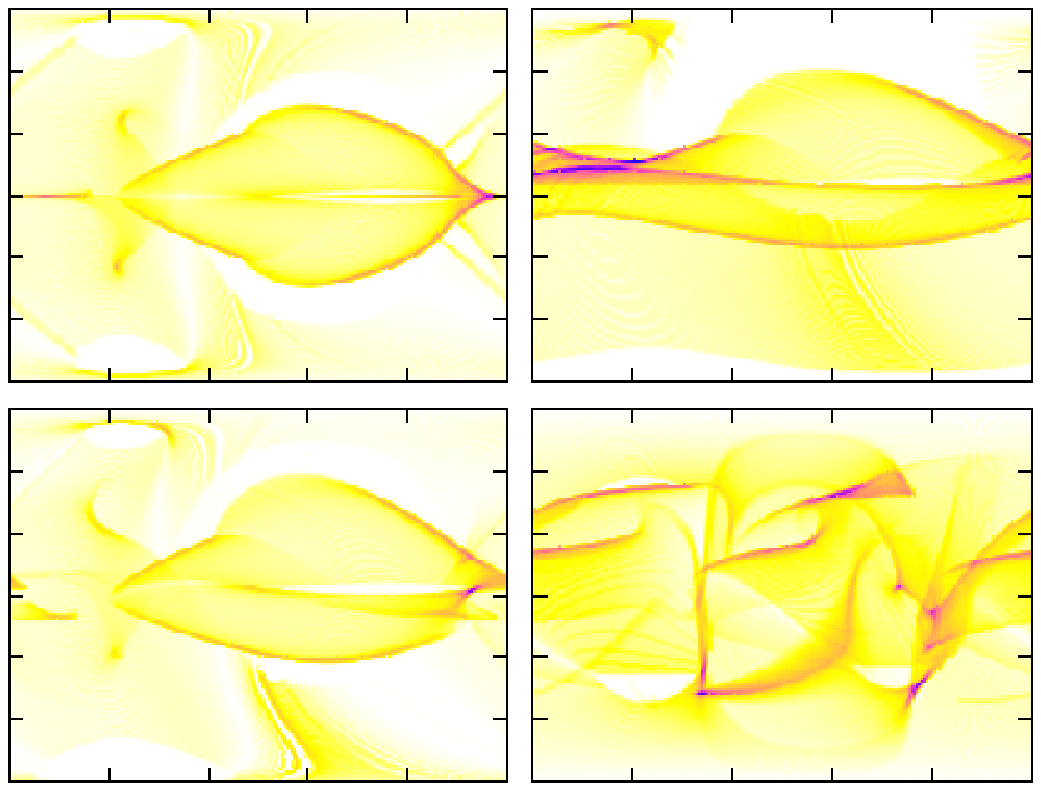}
\caption{Phase diagram depending on the dipole and quadrupole and the inclination angles $ \chi, \chi_1, \chi_2$ for weights $w_{\rm d} = w_{\rm q} = 1$.}
\label{fig:PhaseDiagramDipoleQuadrupole3}
\end{figure*}

The most interesting case uses a linear combination of dipole and quadrupole as done in equation~(\ref{eq:Bdipquad}). The variety of light-curves is shown in Fig.~\ref{fig:PhaseDiagramDipoleQuadrupole2} where the caustics are identified in red-blue-black colour. All the possible light-curves are present, four pulses, three pulse, two pulses, one pulse, and sometimes with bridge emission.
Another sample is shown in Fig.~\ref{fig:PhaseDiagramDipoleQuadrupole3}. A degeneracy between the angles $(\chi,\chi_1,\chi_2)$ shows that different combinations leads to the same phase diagram, modulo some translation in phase or reflection with respect to the equatorial plane. This is the case for $(\chi=90^o,\chi_1=180^o,\chi_2=0^o)$ which is very similar to $(\chi=90^o,\chi_1=0^o,\chi_2=0^o)$ apart from a translation of half a period. The same applies for $(\chi=30^o,\chi_1=180^o,\chi_2=0^o)$ and $(\chi=30^o,\chi_1=0^o,\chi_2=0^o)$ which are reflection symmetric to each other with respect to $\zeta=90^o$ and translated by half a period. This expectation is immediately seen from the chosen expressions for the constants $q_{l,m}$ possessing themselves some (anti-)symmetry properties.

Symmetry considerations can significantly decrease the number of relevant plots to show in a an atlas.

\section{CONCLUSION}
\label{sec:Conclusion}

It is hard to believe that a neutron star only possess a dipolar magnetic field. Although the rotating dipole is the dominant component in the far wave zone, higher order multipoles must be present close to its surface and in its magnetosphere. We solve analytically the time harmonic Maxwell equations in vacuum and the corresponding expression for the Poynting flux taking into account the finite size of the star. We derived useful analytical expressions for the Poynting flux, the braking index and the retarded point multipole fields. We briefly discussed the important implications for magnetic field estimates at the neutron star surface, the radio pulse and high energy emission properties. We showed that already in the vacuum case, expectations and predictions relative to a pure dipole are drastically diverging. For instance, a double peaked light curve separated by approximately half a period can be explained by a geometric configuration deviating significantly from an orthogonal rotator if we accept the presence of a quadrupolar field of the same intensity as the dipole.

We plan to look more deeply into the phase resolved radio polarization characteristics to constrain the field line geometry close to the star. Including our picture in a force-free pulsar magnetosphere would also alleviate the conclusions drawn so far in the literature about pulsar electrodynamics in a cold massless plasma.

\section*{Acknowledgements}

I am grateful to the referee for his help to improve the readability of the paper. This work has been supported by the French National Research Agency (ANR) through the grant No. ANR-13-JS05-0003-01 (project EMPERE). It also benefited from the computational facilities available at Equip@Meso (Universit\'e de Strasbourg).


\appendix

\onecolumn

\section{Comparison with previous work}
\label{app:A}

As a check of our formalism, we compare our approach to the more cumbersome technique used by \cite{2015A&A...573A..51B}. For the axisymmetric multipole magnetic field, if we substitute their definition of the constant~$B_l^0$ by
\begin{equation}
 R^{l+2} \, B_l^0 = - a_{l,0}^B \, \sqrt{\frac{l\,(l+1)\,(2\,l+1)}{4\,\upi}}
\end{equation}
we retrieve the magnetic field given in their equation~(16). For the electric part we get
\begin{subequations}
\begin{align}
  D_r & = \varepsilon_0 \, \Omega \, f_{l,0}^B \, \left[ (l+2) \, \sqrt{\frac{l}{4\,\upi\,(2\,l+1)}} \, \left( \frac{R}{r} \right)^{l+3} \, P_{l+1}^0 - l \, \sqrt{\frac{l\,(l+1)}{4\,\upi\,(2\,l+1)}} \, \left( \frac{R}{r} \right)^{l+1} \, P_{l-1}^0 \right] \nonumber \\
  & = \varepsilon_0 \, \frac{\Omega \, B_l^0 \, R}{2\,l+1} \, \left[ \left( \frac{R}{r} \right)^{l+1} \,l \,  P_{l-1}^0 - \left( \frac{R}{r} \right)^{l+3} \, (l+2) \, P_{l+1}^0 \right]
\end{align}
\end{subequations}
and
\begin{subequations}
\begin{align}
  D_\vartheta & = \varepsilon_0 \, \Omega \, f_{l,0}^B \, \left[ - (l+1) \, \sqrt{\frac{l\,(l+1)}{4\,\upi\,(2\,l+1)}} \, \left( \frac{R}{r} \right)^{l+3} \, \frac{dP_{l+1}^0}{d\vartheta} + \sqrt{\frac{l\,(l+1)}{4\,\upi\,(2\,l+1)}} \, \left( \frac{R}{r} \right)^{l+1} \, \frac{P_{l-1}^0}{d\vartheta} \right] \nonumber \\
  & = \varepsilon_0 \, \frac{\Omega \, B_l^0 \, R}{2\,l+1} \, \left[ \left( \frac{R}{r} \right)^{l+3} \, \frac{dP_{l+1}^0}{d\vartheta} - \left( \frac{R}{r} \right)^{l+1} \, \frac{P_{l-1}^0}{d\vartheta} \right]
\end{align}
\end{subequations}
This is the opposite of the electric field given in equation~(21) of \cite{2015A&A...573A..51B} (apart from the monopolar term that we did not include in our work). Indeed their $E_\vartheta$ component is not consistent with their boundary condition~(19) and recurrence formula~(20). It should be the opposite sign as given above.

For the asymmetric case, to retrieve the expansion into transverse electric (TE) and transverse magnetic (TM) modes, we should make the following identification between our notations and the one used by \cite{2015A&A...573A..51B}
\begin{subequations}
 \begin{align}
 \label{eq:ATE}
  A_{l,m}^{\rm TE} & = - k_m \, a_{l,m}^B \, \sqrt{l\,(l+1)} \, \sqrt{\frac{2\,l+1}{4\,\upi} \, \frac{(l-m)!}{(l+m)!}} \\
 \label{eq:ATM}
  \varepsilon_0 \, A_{l,m}^{\rm TM} & = - k_m \, a_{l,m}^D \, \sqrt{l\,(l+1)} \, \sqrt{\frac{2\,l+1}{4\,\upi} \, \frac{(l-m)!}{(l+m)!}} .
 \end{align}
\end{subequations}
Therefore the TE mode is associated in our picture to the $f_{l,m}^B$ part of expansion in equation~(\ref{eq:Decomposition_HSV_div_0_D})-(\ref{eq:Decomposition_HSV_div_0_B}) and the TM mode is associated to the $f_{l,m}^D$ part of the same expansion. It is straightforward to check that we indeed get the magnetic and electric field components as reported in equations(26)-(29) of \cite{2015A&A...573A..51B}.
The last step concerns the imposition of the boundary conditions on the neutron star surface. Adopting their definition $B_{r;lm}^<$ of the internal field we have
\begin{equation}
 B_{r;lm}^<(R) = - \sqrt{l\,(l+1)} \, \sqrt{\frac{2\,l+1}{4\,\upi} \, \frac{(l-m)!}{(l+m)!}} \, \frac{f_{l,m}^B(R)}{R} .
\end{equation}
Using our definition of $f_{l,m}^B(R)$ from eq.~(\ref{eq:aBlm}) and the identification of the TE modes in eq.(\ref{eq:ATE}) we get
\begin{equation}
 A_{l,m}^{\rm TE} = B_{r;lm}^<(R) \, \frac{k_m\,R}{h_l^{(1)}(k_m\,R)}
\end{equation}
which is nothing but equation~(30) of \cite{2015A&A...573A..51B}. For the TM modes, using the identification~eq.~(\ref{eq:ATM}) and our boundary conditions eq.~(\ref{eq:aDlm}) we get
\begin{subequations}
 \begin{align}
  A_{l+1,m}^{\rm TM} & = A_{l,m}^{\rm TE} \, \frac{(l-m+1)\,(l+2)}{(2\,l+1)\,m} \, \frac{k_m\,R \, h_l^{(1)}(k_m\,R)}{|\partial_r ( r \, h_{l+1}^{(1)}(k_m\,r))|_R} \\
  A_{l-1,m}^{\rm TM} & = - A_{l,m}^{\rm TE} \, \frac{(l+m)\,(l-1)}{(2\,l+1)\,m} \, \frac{k_m\,R \, h_l^{(1)}(k_m\,R)}{|\partial_r ( r \, h_{l-1}^{(1)}(k_m\,r))|_R} .
 \end{align}
\end{subequations}
These are identical respectively to equation~(32) and equation~(33) of \cite{2015A&A...573A..51B}. Consequently, we demonstrated that our formalism employing vector spherical harmonics gives exactly the same results as a decomposition into TE and TM modes.

\section{Retarded point multipoles}
\label{app:B}

\subsection{Retarded point dipole}

Boundary conditions on the neutron star reflect in the vacuum solution, even very far away in the wave zone, through additional $l=2$ terms in the electromagnetic field. Their effect is dominant for extremely rapidly rotating stars, $a=R/\rlight\lesssim1$. As neutron stars are observed to rotate at most with a period of 1.6~ms, the largest ratio of $R/\rlight$ corresponds roughly to $0.1$. The Deutsch solution can then be expanded in a rapidly converging series of $R/\rlight$. For practical purposes, it is a very good approximation, even to the lowest order. To get this approximation, we compute the solution in the limit of vanishing stellar radius by setting $R\to0$ in the Deutsch solution given for instance in full length in \cite{2012MNRAS.424..605P}. Nevertheless we assume a constant magnetic moment $\mu=B\,R^3$ in this limit. We obtain the following expressions for the component of the magnetic field in spherical coordinates
\begin{subequations}
\label{eq:DipoleRetarde}
 \begin{align}
  B_r & = \frac{2\,B\,R^3}{r^3} \, \left[ \cos\chi\,\cos\vartheta + \sin\chi\,\sin \vartheta \, ( \cos\psi + k\,r\,\sin\psi ) \right] \\
  B_\vartheta & = \frac{B\,R^3}{r^3} \, \left[ \cos\chi\,\sin\vartheta + \sin\chi\,\cos \vartheta \, \left\{ (k^2\,r^2-1) \, \cos\psi - k\,r\,\sin\psi \right\} \right] \\
  B_\varphi & = - \frac{B\,R^3}{r^3} \, \sin\chi\, \left[ k\,r\, \cos\psi + (k^2\,r^2-1)\,\sin\psi \right] ,
 \end{align}
\end{subequations}
with the instantaneous phase given by $\psi=k\,r-\Omega\,t+\varphi$. We draw the attention of the reader not to confuse the above retarded point dipole solution~(\ref{eq:DipoleRetarde}) with the exact Deutsch field given in equation~(\ref{eq:Deutsch}). In the point dipole solution, the quadrupole terms induced by the corotating quadrupole electric field inside the star are missing because the boundary conditions are irrelevant for a point dipole (so no $\cos 2\vartheta$ term for instance). In general the retarded {\it point} multipole is not exactly the same as the {\it finte size} multipole field, the latter taking into account boundary conditions on a sphere of finite size. In a point multipole description, there is no need to impose boundary conditions, therefore they are no higher multipolar electric field generated by the rotating magnetic field because no charge redistribution in a perfectly conducting sphere. Nevertheless the point dipole approximation is often used to model pulsed emission from neutron star magnetospheres. We found it useful to give simple analytical expressions for any retarded point multipole in the limit of vanishing radius of the sphere. It is computationally more efficient without losing accuracy (to second order in $R/\rlight$). See for instance \cite{2004ApJ...606.1125D} for the retarded point dipole in cartesian coordinates and \cite{2014MNRAS.437..262M} on how to derive this simple point dipole solution with vector algebra. These components are used to compute magnetic field lines in pulsar dipolar magnetospheres to get the shape of the polar caps, see \S\ref{sec:PolarCap} and from which pulsed emission is supposed to come from, see \S\ref{sec:Emission}. The same procedure as before can be reproduced for higher order multipoles. We show three more examples concerning the quadrupole, the hexapole and the octopole in the next paragraphs.

\subsection{Retarded point quadrupole}

Expanding the exact vacuum solution to lowest order while keeping the magnetic quadrupole moment constant, we can derive the components of the magnetic field. In this case, we show separately the contribution of each mode~$m$. Note here again that the expressions given below apply only to a point quadrupole, without specifying boundary conditions on the neutron star because its size is zero in this limit. The axisymmetric mode $m=0$ has components with weight $\cos \chi_1 $ such that
\begin{subequations}
 \begin{align}
  B_r & = - \frac{B\,R^4}{r^4} \, \frac{1}{2} \sqrt{\frac{5}{2}} (1 + 3 \cos 2 \vartheta ) \\
  B_\vartheta & = - \frac{B\,R^4}{r^4} \, \sqrt{\frac{5}{2}} \sin 2 \vartheta \\
  B_\varphi & = 0
 \end{align}
\end{subequations}
The first asymmetric mode, $m=1$, has components with weight $\sin \chi_1 \cos \chi_2$ such that
\begin{subequations}
 \begin{align}
 B_r & = \frac{B\,R^4}{r^4} \, \sqrt{\frac{5}{6}} \sin 2 \vartheta  ((3-k^2 r^2) \cos \psi + 3 k r \sin \psi ) \\
 B_\vartheta & = \frac{B\,R^4}{r^4} \, \frac{1}{3 \sqrt{30}} \, ( 3 \cos \psi  (5 \cos 2 \vartheta  (k^2 r^2-2)+3 k^2 r^2) + \nonumber \\
 & k r \sin \psi  (5 \cos 2 \vartheta  (k^2 r^2-6)+9 k^2 r^2) \\
 B_\varphi & = \frac{B\,R^4}{r^4} \, \frac{1}{3} \sqrt{\frac{2}{15}} \cos \vartheta  (3
   (5-4 k^2 r^2) \sin \psi +k r (7 k^2 r^2-15) \cos \psi )
 \end{align}
\end{subequations}
The last mode $m=2$ has components with weight $\sin \chi_1 \sin \chi_2$ and are given by
\begin{subequations}
 \begin{align}
 B_r & = \frac{B\,R^4}{r^4} \, \sqrt{\frac{5}{6}} \sin ^2\vartheta  ((4 k^2 r^2 - 3 ) \cos 2 \psi - 6 k r \sin 2 \psi ) \\
 B_\vartheta & = \frac{B\,R^4}{r^4} \, \frac{1}{3} \sqrt{\frac{5}{6}} \sin 2 \vartheta  (2 k r (3 - 2 k^2 r^2 ) \sin 2 \psi + (3 - 6 k^2 r^2 ) \cos 2 \psi ) \\
 B_\varphi & = \frac{B\,R^4}{r^4} \, \frac{1}{3} \sqrt{\frac{10}{3}} \sin \vartheta  (3 (2 k^2 r^2 - 1 ) \sin 2 \psi + 2 k r () 3 - 2 k^2 r^2 ) \cos 2 \psi )
 \end{align}
\end{subequations}
These expressions together with the retarded point dipole are used to compute the polar cap shape as well as the high-energy pulsed emission in the multipole-caustic model described in \S\ref{sec:PolarCap} and \S\ref{sec:Emission}.

\subsection{Retarded point hexapole}

By expansion of the exact vacuum solution to lowest order, keeping the magnetic hexapole moment constant, we show separately the component of each mode~$m$.  Note here again that the expressions given below apply only to a point hexapole, without specifying boundary conditions on the neutron star because its size is zero in this limit. The axisymmetric mode $m=0$ has components with weight $\cos \chi_1$ such that
\begin{subequations}
 \begin{align}
  B_r & = - \frac{B\,R^5}{r^5} \, \frac{1}{2} \sqrt{\frac{7}{2}} ( 3 \cos\vartheta + 5 \cos 3 \vartheta ) \\
 B_\vartheta & = - \frac{B\,R^5}{r^5} \, \frac{3}{8} \sqrt{\frac{7}{2}} ( \sin\vartheta + 5 \sin 3 \vartheta ) \\
 B_\varphi & = 0.
 \end{align}
\end{subequations}
The first asymmetric mode $m=1$ has components with weight $\sin \chi_1 \cos \chi_2 \cos \psi$ given by
\begin{subequations}
 \begin{align}
  B_r & = - \frac{B\,R^5}{r^5} \, \frac{1}{10} \sqrt{21} \sin \vartheta  (5 \cos 2 \vartheta +3) (2 k^2 r^2-5)\\
 B_\vartheta & = - \frac{B\,R^5}{r^5} \, \frac{\cos \vartheta  (37 k^4 r^4+183 k^2 r^2+35 \cos 2 \vartheta  (k^4 r^4-21 k^2 r^2+45)-735)}{40 \sqrt{21}} \\
 B_\varphi & = \frac{B\,R^5}{r^5} \, \frac{k r (5 \cos 2 \vartheta  (46 k^2 r^2-105)+42 k^2 r^2-315)}{40 \sqrt{21}}.
 \end{align}
\end{subequations}
and components with weight $\sin \chi_1 \cos \chi_2 \sin \psi$ given by
\begin{subequations}
 \begin{align}
  B_r & = - \frac{B\,R^5}{r^5} \, \frac{1}{10} \sqrt{\frac{7}{3}} k r \sin \vartheta  (5 \cos 2 \vartheta +3) (k^2 r^2-15) \\
 B_\vartheta & = \frac{B\,R^5}{r^5} \, \frac{k r \cos \vartheta  (105 \cos 2 \vartheta  (2 k^2 r^2-15)+62 k^2 r^2+735)}{40 \sqrt{21}} \\
 B_\varphi & = \frac{B\,R^5}{r^5} \, \frac{5
   \cos 2 \vartheta  (13 k^4 r^4-81 k^2 r^2+105)+7 (k^4 r^4-21 k^2 r^2+45)}{40 \sqrt{21}}.
 \end{align}
\end{subequations}
The mode $m=2$ has components with weight $\sin \chi_1 \sin \chi_2 \cos \chi_3 \cos2\psi$
\begin{subequations}
 \begin{align}
   B_r & = \frac{B\,R^5}{r^5} \, \sqrt{\frac{42}{5}} \sin ^2\vartheta  \cos \vartheta  (8 k^2 r^2-5) \\
  B_\vartheta & = \frac{B\,R^5}{r^5} \, \frac{\sin \vartheta  (7 \cos
   2 \vartheta  (16 k^4 r^4-84 k^2 r^2+45)+3 (48 k^4 r^4-92 k^2 r^2+35))}{4
   \sqrt{210}} \\
  B_\varphi & = -\frac{B\,R^5}{r^5} \, \frac{k r \sin 2 \vartheta  (76 k^2 r^2-105)}{\sqrt{210}}.
 \end{align}
\end{subequations}
and components with weight $\sin \chi_1 \sin \chi_2 \cos \chi_3 \sin2\psi$
\begin{subequations}
 \begin{align}
   B_r & = \frac{B\,R^5}{r^5} \, \sqrt{\frac{14}{15}} k r \sin \vartheta  \sin 2 \vartheta  (4 k^2 r^2-15) \\
  B_\vartheta & = - \frac{B\,R^5}{r^5} \, \frac{k r \sin \vartheta 
   (21 \cos 2 \vartheta  (8 k^2 r^2-15)+136 k^2 r^2-105)}{2 \sqrt{210}} \\
  B_\varphi & = \frac{B\,R^5}{r^5} \, 18 \sqrt{\frac{6}{35}}
   k^2 r^2 \sin 2 \vartheta .
 \end{align}
\end{subequations}
Finally, the last mode $m=3$ has components with weight $\sin \chi_1 \sin \chi_2 \sin \chi_3 \cos3\psi$ is
\begin{subequations}
 \begin{align}
  B_r & = - \frac{B\,R^5}{r^5} \, \sqrt{\frac{7}{5}} \sin ^3\vartheta  (18 k^2 r^2-5) \\
 B_\vartheta & = - \frac{B\,R^5}{r^5} \, \frac{3}{4} \sqrt{\frac{7}{5}} \sin ^2\vartheta 
   \cos \vartheta  (9 k^4 r^4-21 k^2 r^2+5) \\
 B_\varphi & = \frac{B\,R^5}{r^5} \, \frac{9}{4} \sqrt{\frac{7}{5}} k r \sin ^2\vartheta  (6 k^2
   r^2-5)
 \end{align}
\end{subequations}
and components with weight $\sin \chi_1 \sin \chi_2 \sin \chi_3 \sin3\psi$
\begin{subequations}
 \begin{align}
  B_r & = - \frac{B\,R^5}{r^5} \, 3 \sqrt{\frac{7}{5}} k r \sin ^3\vartheta  (3 k^2 r^2-5)\\
 B_\vartheta & =  \frac{B\,R^5}{r^5} \, \frac{9}{4} \sqrt{\frac{7}{5}} k r \sin
   ^2\vartheta  \cos \vartheta  (6 k^2 r^2-5) \\
 B_\varphi & = \frac{B\,R^5}{r^5} \,\frac{3}{4} \sqrt{\frac{7}{5}} \sin ^2\vartheta  (9 k^4
   r^4-21 k^2 r^2+5)
 \end{align}
\end{subequations}

\subsection{The retarded point octopole}

Performing the lowest order expansion of this octopole, we get the retarded point octopole which is splitting into its various azimuthal modes~$m$. Note here again that the expressions given below apply only to a point octopole, without specifying boundary conditions on the neutron star because its size is zero in this limit. Explicitly, the mode $m=0$ has components with weight $\cos \chi_1$ and 
\begin{subequations}
 \begin{align}
  B_r & = - \frac{B\,R^6}{r^6} \, \frac{1}{32} \sqrt{\frac{15}{2}} (20 \cos 2 \vartheta +35 \cos 4 \vartheta +9) \\
  B_\vartheta & = - \frac{B\,R^6}{r^6} \, \frac{1}{8} \sqrt{\frac{15}{2}} (2 \sin 2 \vartheta +7 \sin 4 \vartheta ) \\
  B_\varphi & = 0.
 \end{align}
\end{subequations}
The mode $m=1$ has components with weight $\sin \chi_1 \cos \chi_2 \cos\psi$ 
\begin{subequations}
 \begin{align}
  B_r & = \frac{B\,R^6}{r^6} \, \frac{(2 \sin 2 \vartheta +7 \sin 4 \vartheta ) (k^4 r^4-45 k^2 r^2+105)}{112 \sqrt{3}} \\
  B_\vartheta & = - \frac{B\,R^6}{r^6} \, \frac{4 \cos
   2 \vartheta  (19 k^4 r^4-73 k^2 r^2+63)+21 (k^2 r^2 (2 k^2 r^2-5)+\cos 4 \vartheta 
   (2 k^4 r^4-39 k^2 r^2+84))}{336 \sqrt{3}} \\
  B_\varphi & = - \frac{B\,R^6}{r^6} \, \frac{k r \cos \vartheta  (-59 k^4 r^4+795 k^2
   r^2+21 \cos 2 \vartheta  (7 k^4 r^4-135 k^2 r^2+315)+945)}{1260 \sqrt{3}}.
 \end{align}
\end{subequations}
and components with weight $\sin \chi_1 \cos \chi_2 \sin\psi$
\begin{subequations}
 \begin{align}
  B_r & = - \frac{B\,R^6}{r^6} \, \frac{5 k r (2 \sin 2 \vartheta +7 \sin 4 \vartheta ) (2 k^2 r^2-21)}{112 \sqrt{3}} \\
  B_\vartheta & = - \frac{B\,R^6}{r^6} \, \frac{k r}{5040 \sqrt{3}} \, ( 4 \cos 2 \vartheta (46 k^4 r^4-780 k^2 r^2+945)+ \nonumber \\
  & 21 (5 k^2 r^2 (k^2 r^2-15) + 3 \cos 4 \vartheta  (k^4 r^4-55 k^2 r^2+420)) ) \\
  B_\varphi & = \frac{B\,R^6}{r^6} \,\frac{\cos \vartheta  (-23 k^4 r^4+32 k^2 r^2+21 \cos 2 \vartheta  (3 k^4 r^4-16 k^2 r^2+21)+63)}{84 \sqrt{3}} .
 \end{align}
\end{subequations}
The mode $m=2$ has components with weight $\sin \chi_1 \sin \chi_2 \, \cos \chi_3 \, \cos2\psi$ 
\begin{subequations}
 \begin{align}
  B_r & = - \frac{B\,R^6}{r^6} \, \frac{\sin ^2\vartheta  (7 \cos 2 \vartheta +5) (16 k^4 r^4-180 k^2 r^2+105)}{28 \sqrt{6}} \\
  B_\vartheta & = \frac{B\,R^6}{r^6} \,\frac{\sin
   2 \vartheta  (88 k^4 r^4+47 k^2 r^2+21 \cos 2 \vartheta  (8 k^4 r^4-39 k^2 r^2+21)-63)}{42
   \sqrt{6}} \\
  B_\varphi & = \frac{B\,R^6}{r^6} \, \frac{k r \sin \vartheta  (21 \cos 2 \vartheta  (32 k^4 r^4-240 k^2 r^2+315)+5 (64 k^4
   r^4-600 k^2 r^2+945))}{315 \sqrt{6}}.
 \end{align}
\end{subequations}
and components with weight $\sin \chi_1 \sin \chi_2 \, \cos \chi_3 \, \sin 2\psi$
\begin{subequations}
 \begin{align}
  B_r & = \frac{B\,R^6}{r^6} \, \frac{5 k r \sin ^2\vartheta  (7 \cos 2 \vartheta +5) (8 k^2 r^2-21)}{14 \sqrt{6}} \\
  B_\vartheta & = \frac{B\,R^6}{r^6} \, \frac{k r \sin 2 \vartheta  (21 \cos 2 \vartheta  (4 k^4 r^4-55 k^2 r^2+105)-5 (37 k^2 r^2+63))}{105
   \sqrt{6}} \\
  B_\varphi & = -\frac{B\,R^6}{r^6} \,\frac{\sin \vartheta  (176 k^4 r^4-620 k^2 r^2+21 \cos 2 \vartheta  (16 k^4 r^4-44 k^2 r^2+21)+315)}{42 \sqrt{6}}.
 \end{align}
\end{subequations}
The mode $m=3$ has components with weight $\sin \chi_1 \sin \chi_2 \sin \chi_3 \, \cos \chi_4 \, \cos 3\psi$ 
\begin{subequations}
 \begin{align}
 B_r & = \frac{B\,R^6}{r^6} \, \frac{1}{2} \sqrt{\frac{3}{7}} \sin ^3\vartheta  \cos \vartheta  (27 k^4 r^4-135 k^2
   r^2+35) \\
  B_\vartheta & = - \frac{B\,R^6}{r^6} \, \frac{\sin ^2\vartheta  (144 k^4 r^4-193 k^2 r^2+3 \cos 2 \vartheta  (54 k^4 r^4-117 k^2
   r^2+28)+42)}{4 \sqrt{21}} \\
  B_\varphi & = - \frac{B\,R^6}{r^6} \, \frac{1}{10} \sqrt{\frac{3}{7}} k r \sin ^2\vartheta  \cos \vartheta  (87
   k^4 r^4-415 k^2 r^2+315)
 \end{align}
\end{subequations}
and components with weight $\sin \chi_1 \sin \chi_2 \sin \chi_3 \, \cos \chi_4 \, \sin 3\psi$
\begin{subequations}
 \begin{align}
 B_r & = - \frac{B\,R^6}{r^6} \, \frac{15}{2} \sqrt{\frac{3}{7}} k r \sin ^3\vartheta  \cos \vartheta  (6 k^2 r^2-7) \\
  B_\vartheta & = - \frac{B\,R^6}{r^6} \,\frac{1}{20}
   \sqrt{\frac{3}{7}} k r \sin ^2\vartheta  (93 k^4 r^4-335 k^2 r^2+\cos 2 \vartheta  (81 k^4 r^4-495 k^2
   r^2+420)+210) \\
  B_\varphi & = \frac{B\,R^6}{r^6} \, \frac{\sin ^2\vartheta  \cos \vartheta  (153 k^4 r^4-272 k^2 r^2+63)}{2 \sqrt{21}}
 \end{align}
\end{subequations}
The mode $m=4$ has components with weight $\sin \chi_1 \sin \chi_2 \sin \chi_3 \, \sin \chi_4 \, \cos 4\psi$
\begin{subequations}
 \begin{align}
  B_r & = - \frac{B\,R^6}{r^6} \,\frac{\sin ^4\vartheta  (256 k^4 r^4-720 k^2 r^2+105)}{4 \sqrt{42}}  \\
  B_\vartheta & = \frac{B\,R^6}{r^6} \, \frac{\sin ^3\vartheta  \cos
   \vartheta  (128 k^4 r^4-156 k^2 r^2+21)}{\sqrt{42}} \\
  B_\varphi & = \frac{B\,R^6}{r^6} \, \frac{2}{5} \sqrt{\frac{2}{21}} k r \sin ^3\vartheta 
   (64 k^4 r^4-220 k^2 r^2+105)
 \end{align}
\end{subequations}
and components with weight $\sin \chi_1 \sin \chi_2 \sin \chi_3 \, \sin \chi_4 \, \sin 4\psi$
\begin{subequations}
 \begin{align}
  B_r & = \frac{B\,R^6}{r^6} \,\frac{5 k r \sin ^4\vartheta  (32 k^2 r^2-21)}{\sqrt{42}}  \\
  B_\vartheta & = \frac{B\,R^6}{r^6} \, \frac{2}{5} \sqrt{\frac{2}{21}} k r \sin
   ^3\vartheta  \cos \vartheta  (64 k^4 r^4-220 k^2 r^2+105) \\
  B_\varphi & = - \frac{B\,R^6}{r^6} \,\frac{\sin ^3\vartheta  (128 k^4 r^4-156
   k^2 r^2+21)}{\sqrt{42}}
 \end{align}
\end{subequations}

\label{lastpage}

\end{document}